\documentclass[12pt]{article}
\usepackage{graphicx}
\usepackage{amsmath}
\usepackage{longtable}

\begin{document}

\makeatletter
\renewcommand*{\@cite}[2]{{#2}}
\renewcommand*{\@biblabel}[1]{#1.\hfill}
\makeatother

\title{Interstellar Extinction}
\author{G.~A.~Gontcharov\thanks{E-mail: georgegontcharov@yahoo.com}}

\maketitle

Pulkovo Astronomical Observatory, Russian Academy of Sciences, Pul\-kov\-skoe sh. 65, St. Petersburg, 196140 Russia


This review describes our current understanding of interstellar extinction. This differ substantially from the ideas of the 20th century.
With infrared surveys of hundreds of millions of stars over the entire sky, such as 2MASS, SPITZER-IRAC, and WISE, we have looked at the densest
and most rarefied regions of the interstellar medium at distances of a few kpc from the sun. Observations at infrared and microwave wavelengths,
where the bulk of the interstellar dust absorbs and radiates, have brought us closer to an understanding of the distribution of the dust particles
on scales of the Galaxy and the Universe. We are in the midst of a scientific revolution in our understanding of the interstellar medium and dust.
Progress in, and the key results of, this revolution are still difficult to predict. Nevertheless, (a) a physically justified model has been developed for
the spatial distribution of absorbing material over the nearest few kiloparsecs, including the Gould belt as a dust container, which gives an accurate
estimate of the extinction for any object just in terms of its galactic coordinates. It is also clear that (b) the interstellar medium makes up roughly
half the mass of matter in the galactic vicinity of the solar system (the other half is made up of stars, their remnants, and dark matter) and
(c) the interstellar medium and, especially, dust, differ substantially in different regions of space, and deep space cannot be understood by only
studying nearby space.

\newpage

\section*{Changes in the concept of interstellar extinction}

The idea that there is some sort of medium in the space between stars which absorbs starlight has been developed since the time of William Herschel.
Li (2005) has given an historical review of these concepts. Until the end of the twentieth
century, however, the interstellar medium and extinction were regarded only as noise in
the study of stars. In fact, the noise was so insignificant that extinction was invoked by only
Olbers himself for resolving his photometric paradox (``if the number of stars in the Universe
is infinite, then the entire sky should be as bright as the Sun''). It was regarded as significant
only near the galactic plane, where the density of the interstellar medium is high enough
that stars are formed from it. The estimate of the interstellar extinction in the layer near
the galactic plane has hardly changed over the last 170 years: from 1$^m$ per kpc by
Struve (1847) to 1.2$^m$ per kpc by Gontcharov (2012b).
In fact, for a long time it was assumed that even near the galactic plane, most of the
material is in stars. For example, Kulikovskii (1985, p. 146) pointed out that interstellar
matter forms no more than 10\% of the mass of matter in the spiral arms of the Galaxy.

The change in these ideas in the 21st century is one of the reasons for the ongoing revolution
in astronomy. In the Besan\c{c}on model of the galaxy (Robin 2003), which was once popular
but is now obsolete (Gontcharov 2012d), 28\% of the mass was assigned to the interstellar
medium in the galactic neighborhood of the Sun, 59\%, to residual baryonic matter (stars,
brown dwarfs, white dwarfs, neutron stars, and planets), and 13\% to dark matter.
In the new version of this model by Czekaj, et al. (2014), right after Binney and Tremain (2008),
the spatial mass density for the interstellar medium in the Sun's vicinity is taken to be 0.05
(47\%), for residual baryonic matter $0.043-0.049$ (about 44\%), and for dark matter,
$0.01M_{\odot}/pc^3$ (9\%).
McKee, et al. (2015), have analyzed the spatial density of matter in the Sun's vicinity in detail
and obtained a density of 0.041 (42.3\%) for interstellar matter, 0.43 (44.3\%) for residual
baryonic matter, and $0.013M_{\odot}/pc^3$ (13.4\%) for dark matter.
But they noted that hydrogen has been detected far from the galactic plane and that, because
of a spherical spatial distribution relative to the center of the Galaxy, it combines in the
models with dark matter. We can, therefore, see that contemporary estimates of the
interstellar medium contain roughly half the mass of the matter in the part of the Galaxy
near the Sun. This is reasonable if it is assumed that stars are being formed from the medium
in the galactic disk in our time. Given the generally accepted relationship between the masses
of gas and dust, the spatial mass density of dust can be estimated as
$5\cdot10^{-4}M_{\odot}/pc^3$ or, in g$/cm^3$,
\begin{equation}
\label{dens}
3.5\cdot10^{-26}.
\end{equation}

During the 20th century, wavelengths in the range $0.4-1$ $\mu$m were most accessible to
astronomers; in that range the absorption is roughly inversely proportional to wavelength,
i.e.,
\begin{equation}
\label{l1l}
A_{\lambda}\sim1/\lambda.
\end{equation}
In addition, research has been limited predominantly to the region of space next to the Sun,
where the proportionality coefficient $R_V$ between the reddening $E(B-V)$ of a star and
the extinction $A_V$ is a constant, the only universal characteristic of the dusty medium for
the entire space and for the whole range of wavelengths (Kulikovskii 1985, p. 151):
\begin{equation}
\label{avrvebv}
A_V=R_V\cdot E(B-V).
\end{equation}

In the 21st century, however, observations at other wavelengths and at larger distances from
the Sun have revealed a great variety of characteristics of the cosmic dust grains (Draine 2003).
In many wavelength ranges and regions of the Galaxy, Eq. (2) is not satisfied and the ratio of the
total-to-selective extinction $R_V$ is only one of the characteristics of the dusty medium and
it varies in space as well as with wavelength (Voshchinnikov 2012). This has forced scientists to
develop a new branch of science, the physics of cosmic dust. The review by Voshchinnikov (2012)
shows that the observed extinction, reddening, and $R_V$ vary extremely widely for different
directions, distances, and wavelengths and sources of radiation. In recent years these have been
described fairly well by theoretical models with different distribution of dust grains with
respect to size, chemical composition, shape, and other properties.

\section*{The SFD98 map}

Absorbed radiation is reemitted by a dust particle at a longer wavelength. This emission can be
used to evaluate the extinction and the properties of the dust.

In 1998, Schlegel, et al (1998) (referred to as SFD98 in the following) published a map of the
entire sky which could be used to analyze reddening and extinction, although indirectly.
It is a map of the IR emission of dust at $\lambda=100$ $\mu$m as a function of galactic longitude $l$
and latitude $b$. With the dust temperature and the calibrations taken into account, once the
zodiacal light and bright point sources are eliminated, the radiation from the dust should
correspond to the reddening of starlight passing through all the galactic matter along a given
line of sight. This map was compiled using data from the COBE/DIRBE and IRAS/ISSA projects.
SFD98 combines the accuracy of COBE/DIRBE (16\%) and the angular resolution of IRAS/ISSA
(about 6 arcmin). For this reason, SFD98 has been used in more than 9000 studies.
It is the standard for estimating the reddening and extinction of extragalactic objects.
Nevertheless, this is a map of the emission, rather than reddening or extinction as often
erroneously claimed (its data are customarily used with $R_V=3.1$). This map also has
systematic errors which are especially relevant for extragalactic astronomy and cosmology.
\begin{enumerate}
\item Arce and Goodman (1999) have shown that the resolution of the dust temperature map
accompanying SFD98 is only $1.4^{\circ}$ (the resolution of COBE/DIRBE), rather than the 6
arcmin resolution of the main map. This should lead to errors in SFD98 when the temperature
gradient is large. These errors were discovered by Gontcharov (2012b) in a comparison of
SFD98 with Gontcharov (2010) map in dense clouds of the Gould belt.
Similar errors in SFD98 have been found by Schlafly, et al. (2014a), in the vicinity of thin
filamentary structures of the medium when using Pan-STARRS1 multicolor photometry for
more than 500 million stars.
\item Using data from the SDSS catalog (Abazajian, et al. (2009)) for millions of stars,
Berry, et al. (2012), found errors in SFD98 which could be caused by incomplete accounting for
zodiacal light and point IR sources.
\item Arce and Goodman (1999) noted that the emission at $\lambda=100$ $\mu$m was rescaled
in SFD98 into the customary reddening $E(B-V)$ for users using multicolor photometry and
spectroscopy for several hundred elliptical galaxies with small reddening ($E(B-V)<0.3^m$),
since in regions of the sky with large reddening, i.e., near the galactic equator, the galaxies are
not visible. As a result, this calibration of SFD98 is wrong for large reddening: the reddening
in SFD98 is overestimated. This has been confirmed in many papers (references are given in
Cambresy, et al. (2005)).
Schlafly, et al. (2014a), have confirmed this using Pan-STARRS1 photometry for more than 500
million stars and by examining the general problems of rescaling emission maps into reddening
and extinction maps.
\item Yahata, et al. (2007), have compared counts of the number of galaxies from SDSS in 69
areas around the galactic north pole with the SFD98 map. For large reddening, there was a
natural drop in the number of galaxies with increasing reddening, but an unexpected increase
in the number of galaxies with reddening was found for small reddening.
In addition, for small reddening, when it is smaller the galaxies are redder on the average.
But this effect becomes weaker with increasing red shift. In a model of the errors in SFD98,
Yahata, et al. (2007), showed that the systematic errors in calibrating SFD98 with minimal
reddening are at a level of only $0.02^m$, which changes to a total error in the minimum
extinction of $\sigma(A_V)\approx0.1^m$. These authors also suggested that in SFD98 the
underestimate of low reddening is caused by an observed, but not accounted for, emission
from galactic clusters in the far IR (apparently because of intergalactic dust). Wolf (2014) has
compared SFD98 with the full (including ``grey'') extinction at high galactic latitudes based on
observations of roughly 50000 quasars. He confirmed that SFD98 overestimates large reddening
and underestimates low reddening, while neglecting circum- and intergalactic emission
in the far IR.
\end{enumerate}

\begin{figure}
\includegraphics{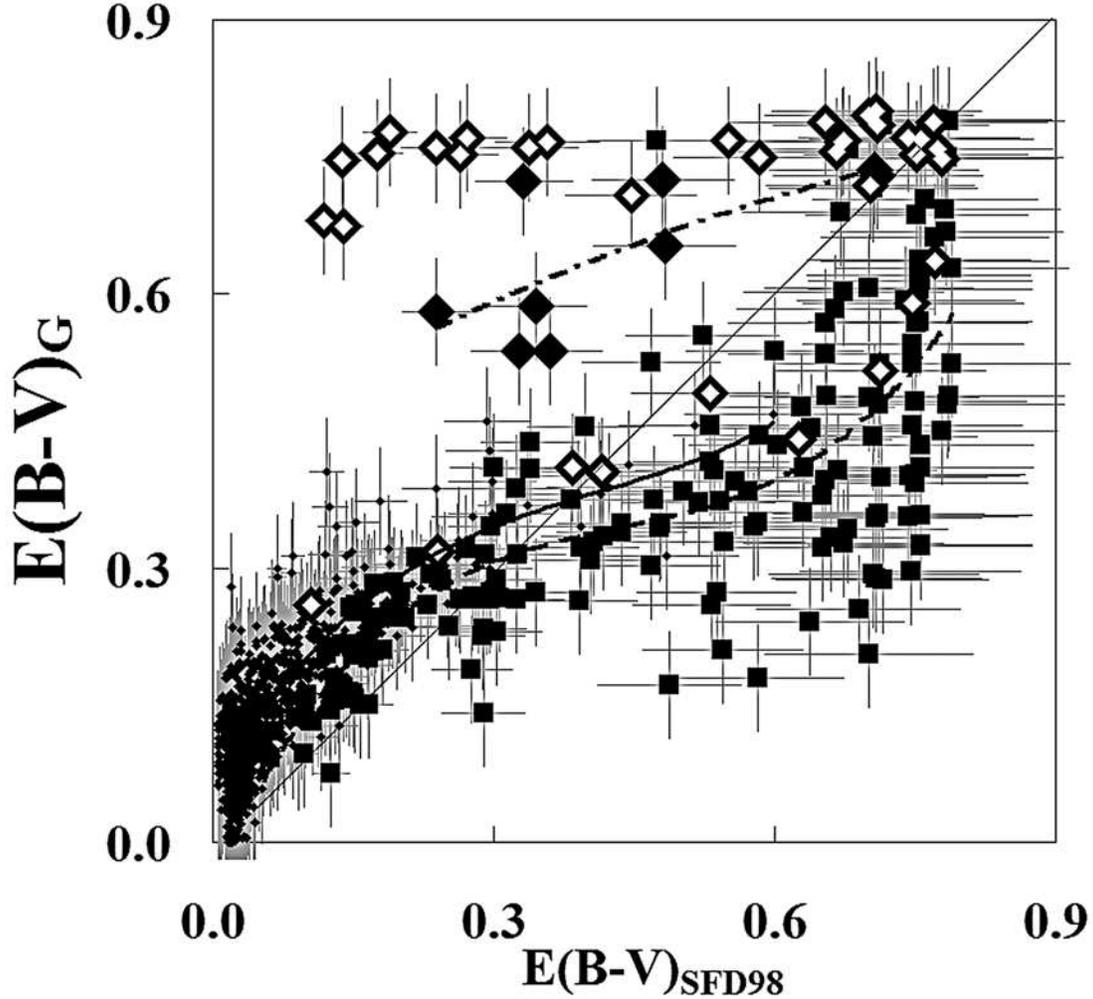}
\caption{The correlation between $E(B-V)_G$ by Gontcharov (2010) and $E(B-V)_{SFD98}$ by
SFD98 for $10^{\circ}\times10^{\circ}$ areas of the sky according to the data of Gontcharov
(2012b). The solid points are data for $|b|>15^{\circ}$ outside of clouds in the Gould
belt and are approximated by the smooth curve.
The large filled diamonds are data for 9 regions with $|b|>15^{\circ}$ containing clouds in the
Gould belt and are approximated by the dot-dash curve. The squares are data for
$|b|<15^{\circ}$ far from the direction of the galactic center and are approximated by the
dashed curve. The large open diamonds are data for the region around the galactic center
($-30^{\circ}<l<+30^{\circ}$, $|b|<15^{\circ}$) which indicate saturation of both maps
with $E(B-V)>0.8^m$. Error bars are shown for all the data.}
\label{fig01}
\end{figure}

As an illustration, Fig. 1 compares the reddening $E(B-V)_G$ according to the Gontcharov (2010)
map for stars at distances of $1.0-1.6$ kpc with $E(B-V)_{SFD98}$ (SFD98, through the entire
Galaxy) for $10^{\circ}\times10^{\circ}$ areas of the sky according to Gontcharov (2012b).
The solid points and the approximate smooth curve are data for $|b|>15^{\circ}$ outside
clouds in the Gould belt.
The large solid diamonds and dot-dash curve are data for 9 regions with $|b|>15^{\circ}$
containing clouds from the Gould belt.
The squares and dashed curve are data for $|b|<15^{\circ}$ far from the direction toward the
galactic center. The large open diamonds are data for the region surrounding the galactic center
($-30^{\circ}<l<+30^{\circ}$, $|b|<15^{\circ}$). Error bars are shown for all these data.
It is clear that near the galactic equator, as expected, the reddening through the entire Galaxy
is considerably greater than the local reddening. But everywhere with $|b|>15^{\circ}$ ,
the differences in the maps agree with their declared high accuracies. It is also clear that the
dot-dash curve is roughly a factor of two higher than the smooth curve and is even above the
dashed curve. Thus, the reddening inside the Gould belt is roughly a factor of two higher than
outside it at the same latitudes and than the reddening near the galactic plane outside the
direction toward the galactic center.

The results are similar for a comparison of SFD98 with the map of Jones, et al. (2011), for high
latitudes at a radius of 2 kpc from the Sun using spectra of more than 9000 class M dwarfs from
the SDSS (Gontcharov, 2012b).
In particular, at high latitudes the average reddening and extinction are
$\overline{E(B-V)}\approx0.06^m$
and $\overline{A_V}\approx0.2^m$, as opposed
to the values from SFD98, which yields
$\overline{E(B-V)}\approx0.03^m$  and $\overline{A_V}\approx0.1^m$.

The underestimated low reddening and overestimated high reddening in the SFD98 map was
confirmed by Gontcharov (2012b) in a comparison of SFD98 with the Gontcharov (2010) and
Jones, et al. (2011) maps with an accuracy that made it possible to express the systematic error
in SFD98 in an analytic form. The smooth curve in Fig. 1 is the least squares polynomial fit
\begin{equation}
\label{sfdcorr}
y=3x^3-3.7x^2+1.8x+0.06,
\end{equation}
where  $y=E(B-V)$ is the true reddening, $x=E(B-V)$ is the reddening according to SFD98,
and it provides a correction to all calculations employing SFD98.

It is difficult to use SFD98 for estimating reddening/extinction within the dust layer of the Galaxy
because of the distances to the structures shown in the map, i.e., this map is two-dimensional.

Unlike SFD98, the three-dimensional emission/reddening/absorption maps, such as those of
Gontcharov (2010) and Schlafly, et al. (2014b), indicate the spatial position of interstellar clouds.
Their distances agree on the whole with the distances obtained by Dame, et al. (1987) and Dutra
and Bica (2002) by comparing the radial velocity of a cloud with a model of galactic rotation
under the assumption that the velocity is determined solely by rotation with no peculiar
velocity or radial motion. But we note that this assumption is questionable and these results
have a low accuracy that appears to be no more than ±500 pc (Gomez (2006)); the distance is
determined only to the leading edge of a cloud, while the extent of the cloud and other clouds
behind it are unseen. The three-dimensional maps determine the position of the clouds much
more accurately, not only the distance to the leading and trailing edge of a cloud but also to all
the hidden clouds along the same line of sight. They confirmed the radial distributions,
relative to the center of the Gould belt (which lies near the sun) of absorbing matter in the
nearest parsec found by Bochkarev and Sitnik (1985) and Straizys, et al. (1999).
In particular, it was confirmed that the Cygnus rift cloud complex extends over a distance from
500 to 2000 pc, is elongated in shape with a size ratio of 1:5, and lies radially relative to the
center of the Gould belt. Thus, it is worth taking note of the assumption in these papers that the
radial orientation (relative to the Sun) of the dust particles and of the entire gigantic dust clouds
may be caused by the special position of the Sun near the epicenter of processes which recently
formed the Local Bubble and the Gould belt. The x-ray emission produced in these processes
could affect the chemical composition and extinction properties of the dust particles, as well as
the galactic magnetic field, which orients the particles.

Despite these shortcomings, the SFD98 map reveals the basic features of the distribution of
absorbing matter in the Galaxy. At high and low latitudes, SFD98 indicates the
reddening/extinction near the Sun, since almost all the absorbing matter for $|b|>15^{\circ}$
(more than 70\% of the sky) lies at a distance of less than 600 pc from the Sun,
and for latitudes $10^{\circ}<|b|<15^{\circ}$ (another 10\% of the sky), closer than 1300 pc
(Gontcharov (2012b)). Note that this attributes a greater significance to studies of reddening
and extinction in the nearest kiloparsecs, especially since most extragalactic objects are
observed in middle and high latitudes. Wolf (2014) has shown that the uncertainty
in the extinction inside the galaxy is one of the main sources of uncertainty in cosmological
parameters derived from type Ia supernovae.

It can be seen in the SFD98 map that the minimum emission/reddening/extinction does not lie
in the galactic poles. In both galactic hemispheres the reddening minimum is double: one pair
of minima ($b\approx+55^{\circ}$, $l\approx160^{\circ}$ and $b\approx-55^{\circ}$,
$l\approx340^{\circ}$) is caused by the orientation of the Gould belt (which contains dust, as
shown below) and the other is caused by the global inclination of the absorbing layer
($b\approx+50^{\circ}$, $l\approx90^{\circ}$ and $b\approx-50^{\circ}$, $l\approx250^{\circ}$).
The second inclination leads to more reddening in the first and second quadrants in the
southern hemisphere and the third and fourth, in the northern hemisphere.

\section*{The model of Arenou, et al. (1992)}

Since the mid-20th century attempts were made to describe extinction with a more or less
simple two- or three-dimensional model or function (one for the sky as a whole or varying from
region to region) that depends, for example, on the galactic coordinates $l$ and $b$ and (to be
desired) the distance $r$. The model differs from the map in that the latter has averaged
observed extinction in cells, while the model approximates all these values by some formula.
In 1992-2009, the best analytic 3D model of interstellar extinction in the nearest kiloparsec was
that of Arenou, et al. (1992), which approximates the average extinction $A_V$ for 199 regions
of the sky with parabolas that depend on distance, i.e.,
$A_V=k_1r+k_2r^2$ , where k1 and k2 are a set of coefficients for a region of the sky.
This model adequately reproduces the observations in the nearest kiloparsec.
It has the shortcoming of lacking any kind of physical justification for the behavior of the
observed variations in $A_V$. As an example, Fig. 2a shows $A_V$ calculated as a
function of $l$ for $r=500$ pc in the bands $+15^{\circ}<b<+30^{\circ}$,
$+5^{\circ}<b<+15^{\circ}$, $-5^{\circ}<b<+5^{\circ}$, $-15^{\circ}<b<-5^{\circ}$,
$-30^{\circ}<b<-15^{\circ}$ using the model of Arenou, et al. (1992).
The vertical lines show the model accuracy declared by the authors.
The common, approximately sinusoidal dependence of $A_V$ on $l$ is plotted in Fig. 1a for
$-5^{\circ}<b<+5^{\circ}$ by a dashed curve corresponding to $0.8+0.5\sin(l+20^{\circ})$.
The Arenou, et al. (1992), model reveals, but does not explain, this dependence, along with the
features of extinction far from the galactic plane. In Fig. 2a it can be seen that for
$+5^{\circ}<b<+30^{\circ}$ the extinction is greater at the longitudes of the galactic center
and for$-30^{\circ}<b<-5^{\circ}$ at the anticenter.

\begin{figure}
\includegraphics{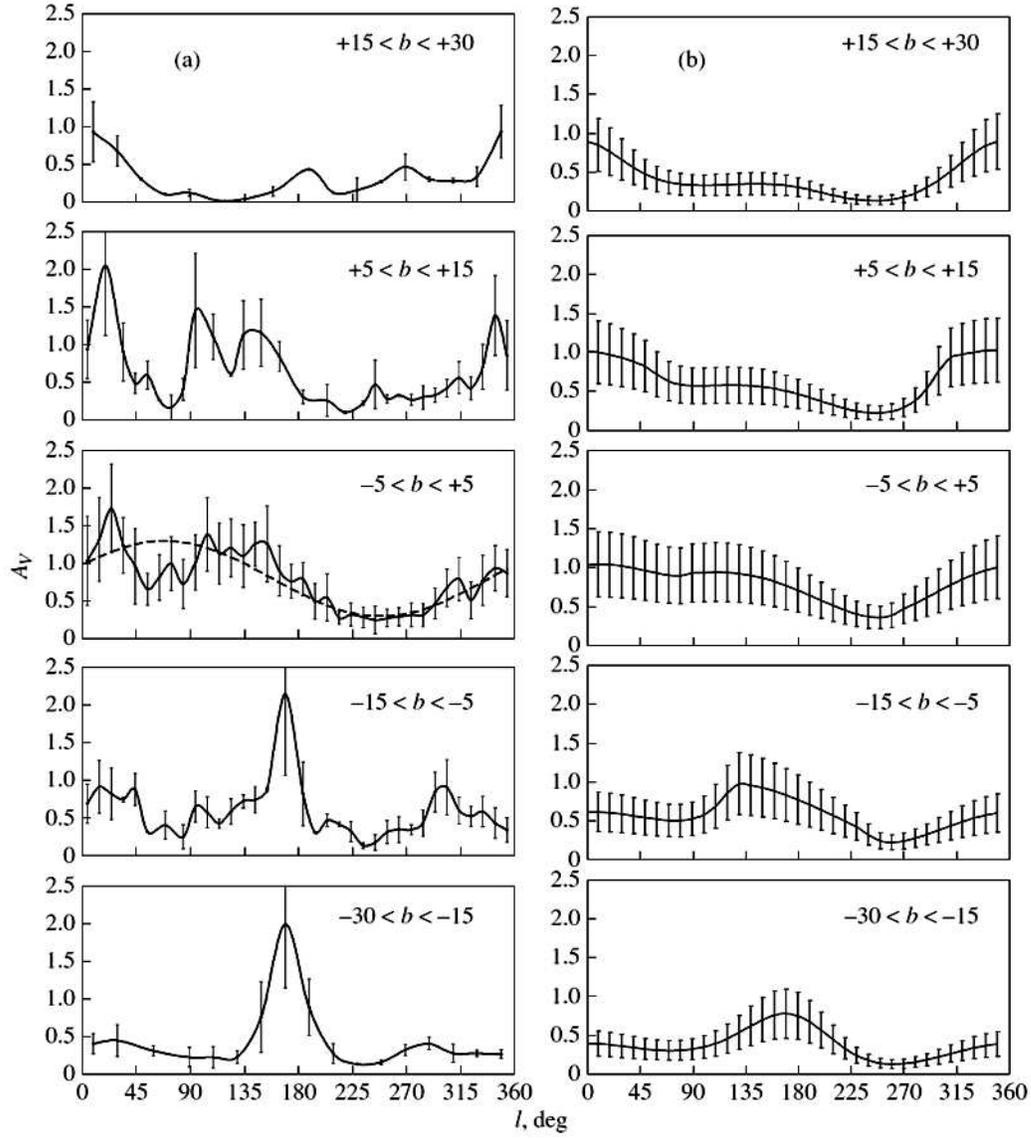}
\caption{$A_V$ as a function of $l$ for $r=500$ pc at different latitudes according to the
models of (a) Arenou, et al. (1992), and (b) Gontcharov (2009). The dashed curve
for $-5^{\circ}<b<+5^{\circ}$ is a plot of $0.8+0.5\sin(l+20^{\circ})$. The vertical lines indicate
the accuracy of the models.}
\label{fig02}
\end{figure}

The accuracy of any model is limited by fluctuations in the extinction from star to star in a given
region of space. For example, for two neighboring stars at distance of 500 pc from us, it is
entirely possible to have spread in extinction of$\sigma(A_V)=0.3^m$ for a typical extinction
$A_V=0.6^m$. These estimates were obtained by Green, et al. (2014), who calculated the
distances, absolute magnitudes, and extinction for roughly a billion stars based on high
precision multicolor photometry data from the Pan-STARRS1 catalog. The analytical model for
extinction in the nearest kiloparsec is more accurate than the Arenou, et al. (1992) is
impossible, but physically better justified models are possible.

The SFD98 and other maps show that for the kiloparsecs closest to the Sun, the main feature of
the distribution of absorbing matter is its concentration in the galactic plane. Another important
feature is the existence of regions with comparatively high extinction far from the galactic plane,
primarily in the Gould belt. The Belt has been described by Gontcharov (2009), Perryman
(2009, pp. 324-328), Gontcharov (2012b), and Bobylev (2014). The Gould belt
contains young stars and associations of stars. Stars are formed here even in our time. The
accompanying interstellar clouds can cause extinction. Taylor, et al. (1996), first pointed out the
Gould belt as a source of dust entering the solar system and Vergely, et al. (1998), were the first
to point out the existence of interstellar extinction in the Gould belt.

\section*{A new model of extinction with dust in the Gould belt}

Figure 3 shows the assumed mutual positions of the two layers of absorbing matter in the
neighborhood of the Sun -- a layer with half thickness $Z_A$ near the equatorial plane of the
Galaxy (the equatorial layer in the following) and a layer in the Gould belt with half thickness
$\zeta_A$ and radius $r'=600$ pc about the Sun. We denote the inclination of the Gould belt to
the galactic plane by $\gamma$. The working coordinate system is defined by the observed
galactic coordinates of a star, $r$, $l$, and $b$, and the Sun is at its center. The displacement of
the principal plane of the equatorial layer relative to the Sun is $Z_0$ and the analogous
displacement for the absorbing layer of the Gould belt is $\zeta_0$.
For comparison with the standard galactic coordinate system, Fig. 3 shows $X'$ and $Y'$, the
axes of a rectangular coordinate system lying in the equatorial layer. The $X'$ axis is parallel to
the direction toward the center of the galaxy and $Y'$ is the direction of galactic rotation.
$\lambda_0$ denotes the turning of the highest point of the Gould belt relative to the
$X'$ axis; it is the angle between the $Y'$ axis and the line of intersection of the layers.

\begin{figure}
\includegraphics{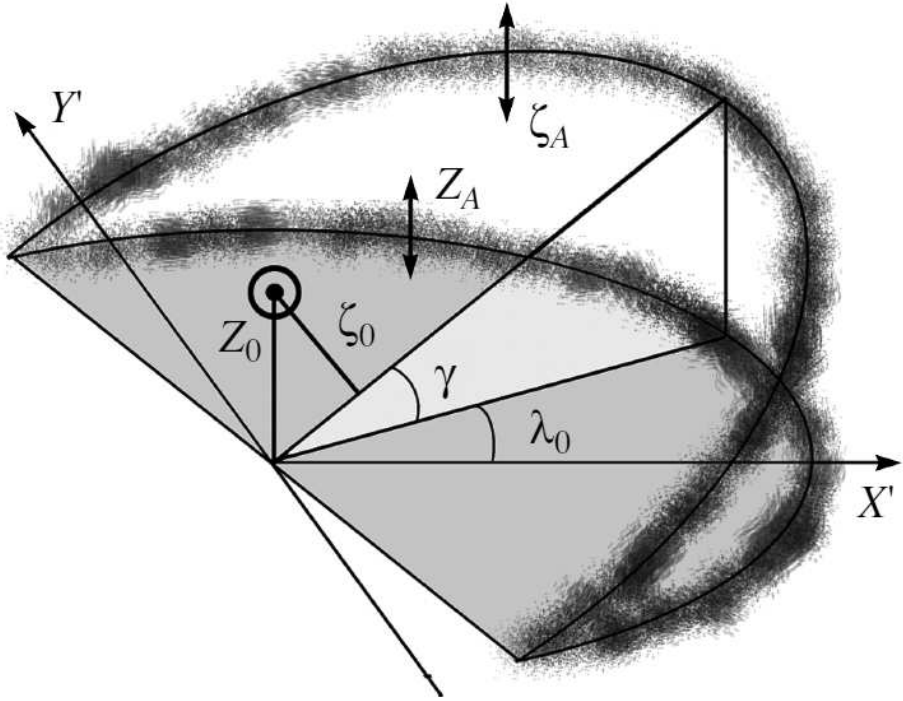}
\caption{The mutual positions of the two layers of dust in the Gontcharov (2009) model.}
\label{fig03}
\end{figure}

The longitude $\lambda$ and latitude $\beta$ of a star relative to the principal plane of the
Gould belt are calculated from the star's galactic coordinates as
\begin{equation}
\label{equ1}
\sin(\beta)=\cos(\gamma)\sin(b)-\sin(\gamma)\cos(b)\cos(l)
\end{equation}
\begin{equation}
\label{equ2}
\tan(\lambda-\lambda_{0})=\cos(b)\sin(l)/(\sin(\gamma)\sin(b)+\cos(\gamma)\cos(b)\cos(l)).
\end{equation}

The observed extinction $A_V$ is approximated as the sum of two functions,
\begin{equation}
\label{aaa}
A_V=A_V(r,l,b)+A_V(r,\lambda,\beta),
\end{equation}
each of which has a barometric dependence (Parenago (1954), p. 265).
The extinction in the equatorial plane is
\begin{equation}
\label{aeq}
A_V(r,l,b)=(A_{0}+A_{1}\sin(l+A_{2}))Z_{A}(1-e^{-r|\sin(b)|/Z_{A}})/|\sin(b)|,
\end{equation}
where $A_0$, $A_1$, $A_2$ are the free term, amplitude, and phase of the extinction in a
sinusoidal dependence on $l$, and the extinction in the Gould belt is
\begin{equation}
\label{ago}
A_V(r,\lambda,\beta)=(\Lambda_{0}+\Lambda_{1}\sin(2\lambda+\Lambda_{2}))\zeta_{A}(1-e^{-r|\sin(\beta)|/\zeta_{A}})/|\sin(\beta)|,
\end{equation}
where where $\Lambda_0$, $\Lambda_1$, and $\Lambda_2$ are the free term, amplitude, and
phase of the extinction in a sinusoidal dependence on $2l$.
The assumption that extinction in the Gould belt has two maxima as a function of the longitude
$\lambda$ was confirmed.
The extinction maxima in the Gould belt are observed near the direction in which the separation
of the Belt from the galactic plane is greatest, i.e., roughly at the longitudes of the center and
anticenter of the Galaxy.

When the displacement of the sun relative to the absorbing layers is taken into account, Eqs. (8)
and (9) transform to
\begin{equation}
\label{aeq2}
A_V(r,l,Z)=(A_{0}+A_{1}\sin(l+A_{2}))r(1-e^{-|Z-Z_{0}|/Z_{A}})Z_{A}/|Z-Z_{0}|
\end{equation}
\begin{equation}
\label{ago2}
A_V(r,\lambda,\zeta)=(\Lambda_{0}+\Lambda_{1}\sin(2\lambda+\Lambda_{2}))\min(r',r)(1-e^{-|\zeta-\zeta_{0}|/\zeta_{A}})\zeta_{A}/|\zeta-\zeta_{0}|,
\end{equation}
where $r'=600$ pc is the limit on the radius of the absorbing layer in the Gould belt (the
extinction beyond the confines of the belt should not be greater than at its edge) and $\zeta$ is
the analog of the distance $Z$ (along the coordinate in the direction of the galactic north pole),
i.e., the displacement of the star in the coordinate system of the Gould belt perpendicular to the
principal plane of the Belt.

As a result, we have a system of Eqs. (7) for each star or cell in space. The left hand sides contain
the observed extinction $A_V$ and the right, a function of the three observed quantities $r$, $l$,
and $b$. Solving these equations yields the values of 12 unknowns:
$\gamma$, $\lambda_{0}$, $Z_{A}$, $\zeta_{A}$, $Z_{0}$, $\zeta_{0}$,
$A_{0}$, $A_{1}$, $A_{2}$, $\Lambda_{0}$, $\Lambda_{1}$, $\Lambda_{2}$
which are chosen so as to minimize the sum of the squares of the discrepancies between the left
and right hand sides of Eqs. (7).

Solutions were obtained using different data on extinction and have been given by Gontcharov
(2009, 2012b).
The most accurate solutions for the extinction in the equatorial layer and in the Gould belt are,
respectively
\begin{equation}
\label{model1}
(1.2+0.3\sin(l+55^{\circ}))r(1-e^{-|Z+0.01|/0.07})0.07/|Z+0.01|,
\end{equation}
\begin{equation}
\label{model2}
(1.2+1.1\sin(2\lambda+130^{\circ}))\min(0.6,r)(1-e^{-|\zeta|/0.05})0.05/|\zeta|,
\end{equation}
where the angle are expressed in degrees and distances in kpc.

The physical justification of the new model is evident in the reliability of the resulting estimates:
the inclination of the Gould belt to the galactic plane is about 19$^{\circ}$, the half thickness of
the dust layers are 70 and 50 pc for the equatorial layer and Gould belt, respectively, the
displacement of the Sun perpendicular to the layers is 10 and 0 pc, respectively, etc.

The other advantages of the new model over the Arenou, et al. (1992), model are simplicity and
continuity:
instead of 199 areas in the sky with individual formulas, we have a single formula; extinction
depends smoothly on the coordinates and does not jump from one area to another.

Today this model is the best way of calculating the extinction $A_V$ for an object if only its
galactic coordinates are known. Near the galactic equator ($|b|<15^{\circ}$) this model is
reliable, at least to the neighboring spiral arms, i.e., out to 2 kpc.
At middle and high latitudes ($|b|>15^{\circ}$) it works out to distances of megaparsecs given
that, as noted above, almost all the absorbing material at these latitudes is within $r<600$ pc.

Therefore, for all objects with $r>600$ pc and $|b|>15^{\circ}$, including most of the
extragalactic objects, this extinction model is not only the best estimate of extinction if only the
coordinates of an object are known, but also gives an acceptable result if the distance of the
object is unknown (then $r=600$ pc in Eq. (13)). For example, for the Andromeda galaxy this
model gives $A_V=0.458^m$ for $r=600$ pc and $A_V=0.459^m$ for $r=750000$ pc.
In any case,
the value according to the new model is substantially higher than the value of $A_V$ usually
assumed for the Andromeda galaxy:
$E(B-V)=0.058^m$ from SFD98 and $A_V=0.18^m$ with $R_V=3.1$.
But, on correcting the value from SFD98 in accordance with Eq. (4), we obtain
$E(B-V)=0.153^m$ and $A_V=0.473$, in good agreement with this estimate according
to the new absorption model.
The reason for the rather high extinction for the Andromeda galaxy is that a region
of the sky near the equatorial plane of the Gould belt is projected onto it.
It provides half the total extinction
($A_V=0.23^m$).
Evidently, up to now this contribution has been underestimated. Because of the discovery of the
importance of the Gould belt as a container of absorbing material, it is necessary to reevaluate
the extinction to extragalactic objects, especially those on which the Belt is projected.

It is important that this extinction model estimates only the extinction in the $V$ band.
It does not yield the extinction in other bands because of spatial variations in the extinction law.

\section*{Three-dimensional maps of extinction}

Straizys' book (1977) is a summary of ideas regarding reddening/extinction up to 1977 and an
indicator of the ongoing scientific revolution in which the requirements and approaches with
which multicolor photometry could be used to set up a sample of stars with a similar spectral
energy distribution (SED). Based on these samples it is possible to study the spatial variations of
reddening, extinction and extinction law. These studies came to fruition in the 21st century with
the publication of catalogs with precise (on a level of $0.01^m$) photometry of millions of stars
over the entire sky in different radiation ranges. The Tycho-2 catalog (H\o g, et al) appeared in
2000, the 2MASS catalog (Skrutskie, et al.) in 2006, and the final version of the WISE catalog
(Wright, et al. (2010)) in 2013.

It was an epochal change. For decades extinction was calculated from reddening (Eq. (3)) under
the assumption that $R_V=3.1$ and reddening, in turn, was calculated as the color excess
between observed color indices of a star and de-reddened color of an unreddened star of the
same spectral class and luminosity class:
\begin{equation}
\label{bv0}
E(B-V)=(B-V)-(B-V)_0.
\end{equation}

But spectral class and luminosity class do not form a unique set of characteristics of the SED of stars. And this is the distribution of interest to us. A unique set is formed by mass, metallicity,
and age. With a small loss of accuracy, they have been transformed into temperature and luminosity or absolute
magnitude and color index, for example, in the PARSEC data base (Bressan, et al. (2012)) of theoretical evolutionary
tracks and isochrones. But calculating spectral and luminosity class using this data is not an unambiguity problem.
Straizys (1977, p. 104) wrote, ``Spectral class and luminosity class are only a crude discrete measure of temperature
and absolute magnitude. In fact, smooth transitions from one spectral class to another and from one luminosity to
another are observed. Thus, a more exact determination of the de-reddened indices of a given star requires
knowledge
of its temperature and absolute magnitude or acceleration of gravity g.''
Even formally, sorting stars in terms of color
index is more fruitful than in terms of spectral class. The subclasses from O5 to M9 for different luminosity classes
represent no more than 200 gradations. And this is all that spectral classification can say about a star. At the same
time, a typical $V=10^m$ star has exact (accuracy on the order of $0.02^m$) photometry from the Tycho-2, 2MASS,
WISE,
and other catalogs. Usually this entails at least 10 photometric bands, i.e., 45 independent color indices. According
to each of these, stars have hundreds of gradations. For example, based on the color index $(u-W2)$ with the $u$
band
from the SDSS catalog and $W2$ from the WISE catalog, the stars cover a range of $1.5^m<(u-W2)<12.0^m$ with
an accuracy
of $0.02^m$, i.e., they have 525 gradations. Therefore, multicolor photometry yields an order of magnitude larger
volume
of information on a star than spectral classification.

The creation of a 3D chart of reddening/extinction based on multicolor photometry assumes that the distance
and approximate SED, i.e., the key astrophysical characteristics, are determined for each star.
All of these quantities must be calculated consistently, usually by solving the following system of equations
(Gontcharov (2012c)):
\begin{equation}
\label{sist}
\left\{
\begin{aligned}
A=f_1(r,l,b),\\
R=f_2(r,l,b),\\
E(m_1-m_2)=A/R,\\
(m_1-m_2)_0=(m_1-m_2)-E(m_1-m_2),\\
M=f_3((m_1-m_2)_0),\\
r=10^{(m-A-M+5)/5},\\
\end{aligned}
\right.
\end{equation}
where f1, f2, and f3 are functions, m1 and m2 are the magnitudes of a star in two bands, M is the absolute
magnitude, $r$ is the distance, $A$ is the extinction, $R$ is the extinction-to-reddening coefficient, and $E$
is the reddening.

Modifications of this method are also widespread. For stars in each class that are close to the Sun we can take
$E(m_1-m_2)=0$ and $(m_1-m_2)=(m_1-m_2)_0$. Then in each distant cell of space, $E(m_1-m_2)=(m_1-m_2)-(m_1-m_2)_0$.
In addition, if the effective temperature, metallicity, and acceleration of gravity are found from an analysis of the spectra
of the stars (as in the SDSS project), then they are used to calculate $(m_1-m_2)_0$ and $M$ before solving the system
(15).

Some examples of this approach: based on 2MASS photometry, Dutra, et al. (2003), selected branch giants from
and constructed a 3D map in the direction of the galactic center. Jones, et al. (2011), have constructed a
3D map for high latitudes within a radius of 2 kpc from the Sun using spectra of more than 9000 class M dwarfs
from SDSS. Here the extinction was determined for each star from an approximation of its spectrum in the
 $0.57<\lambda<0.92$ $\mu$m range of the extinction curve, which depends on $\lambda$
(the variations in $R_V$ were also found, but with an accuracy of only 0.4).
Berry, et al. (2012), have constructed a map based on SDSS and 2MASS photometry for millions of stars.
Chen, et al. (2014), have done so using 2MASS and WISE for 13 million stars in the direction
of the galactic anticenter.
Green, et al. (2014), used Ran STARRS1 data for roughly a billion stars. A 3D map of
reddening (Gontcharov (2010)) and a 3D map of the variation in the extinction law (the coefficient $R_V$) (Gontcharov
(2012a)) compiled by this method have made it possible for Gontcharov (2012b) to construct a 3D map of the
extinction $A_V$ in the nearest kiloparsec with a resolution of 50 pc and an accuracy of $\sigma(A_V)=0.2^m$.

Multiple and peculiar stars which fall into a sample will distort the results. This usually happens if binary
or peculiar stars change the SED only in some bands and a sample is formed from
one set of color indices while the extinction and other characteristics of the sample are calculated from another
set. Some examples: a circumstellar dust shell absorbs short-wavelength emission of a star and reradiates it at
longer wavelengths; there are regions on a star's surface which produce short-wavelength radiation. The usual
solution for this problem is to analyze the SED over as large a set of photometric bands as
possible, eliminating stars with another SED even if it is only in one band. Here all the average color
indices in each cell of space, corrected for reddening, and their standard deviations should be monitored. They should
correspond to theoretical values from, for example, the PARSEC data base. Here iterations can be performed:
reddening is calculated from a ``dirty'' sample and they are used to calculate de-reddened colors, contaminated stars are
rejected, and the reddening is calculated using the ``clear'' sample. This approach has been used by Gontcharov
(2012a). The characteristics of a large number of stars in each cell of space should be averaged. For example,
Gontcharov (2012a) chose the cells so that each contained at least 25 OB stars. Spectral classification for the purpose
of eliminating peculiar stars is usually not effective, since it contains many mistakes.

It is often assumed that $R_V=3.1$ when comparing maps. But comparisons of maps derived from photometry
at substantially different wavelengths shows that there are large discrepancies. The amount is clearly correlated with
the dust temperature (Dutra, et al. (2013); Peek and Graves (2010)). Using an extinction law with $R_V=3.1$ in regions
where it is not satisfied leads to large errors in calculating extinction, distances, absolute magnitudes, and other
characteristics of stars. Here the systematic variations in $R_V$ cause systematic errors in the distances and other
quantities. According to Reis and Corradi (2008), for variations in $R_V$ of $\pm1.5$ from the average, the calculated
photometric distances have errors of 10\%. Thus, spatial variations in the properties of dust and, therefore, the
extinction law cannot be ignored further. Determining the extinction law (or the set of coefficients analogous to
$R_V$ in different radiation bands) is one of the major problems in modern astronomy for the study of the dusty interstellar
medium, reddening, and extinction. $R_V$ and other characteristics of the extinction law reach extreme values, manifest
large variations, and are determined with less accuracy far from the galactic plane and outside the visible spectrum.

The construction of 3D maps has made it clear that low extinction and reddening at high latitudes are not
at all inconsistent with large values of $R_V$ there. In addition, it is now possible to explain the large amount of
conflicting data in the literature on the correlation or anticorrelation of reddening and $R_V$, and thereby, of reddening
and extinction. They are correlated within a thin layer ($|Z|<100$ pc) near the galactic equator, since there the
reddening, $R_V$, and extinction all increase in the direction of the galactic center. Outside this layer they are
anticorrelated: at high and medium latitudes, less reddening corresponds to larger $R_V$. In an extensive study of the
absorption law near the galactic plane, Schlafly, et al. (2016), have found a strong correlation between $R_V$ and emission
in the far IR. This can be explained if the absorption law depends primarily on the size of the dust particles. Then
this anticorrelation shows that fine dust predominates where there are many dust particles and coarse dust, where there
are few. Therefore, it is clear that to some extent the average spatial density of dust is retained in large volumes of
space, despite the variation in the size of the dust particles.

\section*{Variations in the extinction law}

The Weingartner and Draine (2001) extinction law (referred to below as WD2001) can be taken as a standard
extinction law that agrees with observations in some part of the galaxy (but not everywhere). For
$0.4<\lambda<1.2$ $\mu$m and $R_V=3.1$ it agrees with the laws by Draine (2003) and
Indebetouw, et al. (2005), the law derived from Cardelli, et al. (1989), that is assumed in the PARSEC data base,
and others. The WD2001 law is plotted in Fig. 8 for $R_V=3.1$ and $R_V=5.5$ as the lower and upper grey dashed
curves, respectively.

Multicolor photometry can be used to determine the extinction law. In fact, the relationship between two or
more color indices of a star is a characteristic of its SED. For groups of stars with roughly the same initial SED,
a large change in the SED owing to extinction with low reddening corresponds to large values of $R_V$, and vice versa.
Straizys (1977, pp. 39-40) refers to the corresponding method for determining the extinction law as the method of the
extrapolation of extinction law, Berdnikov, et al. (1966), as the method of color differences, Zasowski, et al. (2009), as the
color index ratio method, and Majewski, et al., (2011), as the Raleigh-Jeans color excess method.
This method has been modified
slightly by many authors. It was first used by Johnson and Borgman (1963), who discovered large deviations in $R_V$
from a value of 3.1 with longitude, and found a minimum near $l\approx110^{\circ}$.
(Large deviations in $R_V$ from 3.1 were also reported in the classical paper of Johnson (1965).)

Wegner (2003) has shown that at present this is the only direct method for determining the spatial variations
in the absorption law employing photometry of individual stars, rather than, say, clusters. Attempts to find an
alternative have been made, for example, by finding correlations of $R_V$ with other quantities, such as the
wavelength of maximum polarization of the light or with the characteristic local extinction peak near a wavelength
of 0.2175 $\mu$m. But, as Voshchinnikov (2012) has shown, this yielded questionable results.

Gontcharov (2012a) has shown that using the method of the extrapolation of extinction law requires a complete
sample of high-luminosity stars with roughly the same spectral energy distribution, uniformly distributed over a large
region of space. Here the reddening of these stars must be substantial in each of the spatial cells being considered.
More precisely, the average reddening in a cell must be greater than the natural scatter in the color indices of
unreddened stars and, also, greater than the error in the color indices owing to errors in the original photometry. In
fact, this method has been fruitful for photometric accuracies at a level no worse than $0.01^m$.

\begin{figure}
\includegraphics{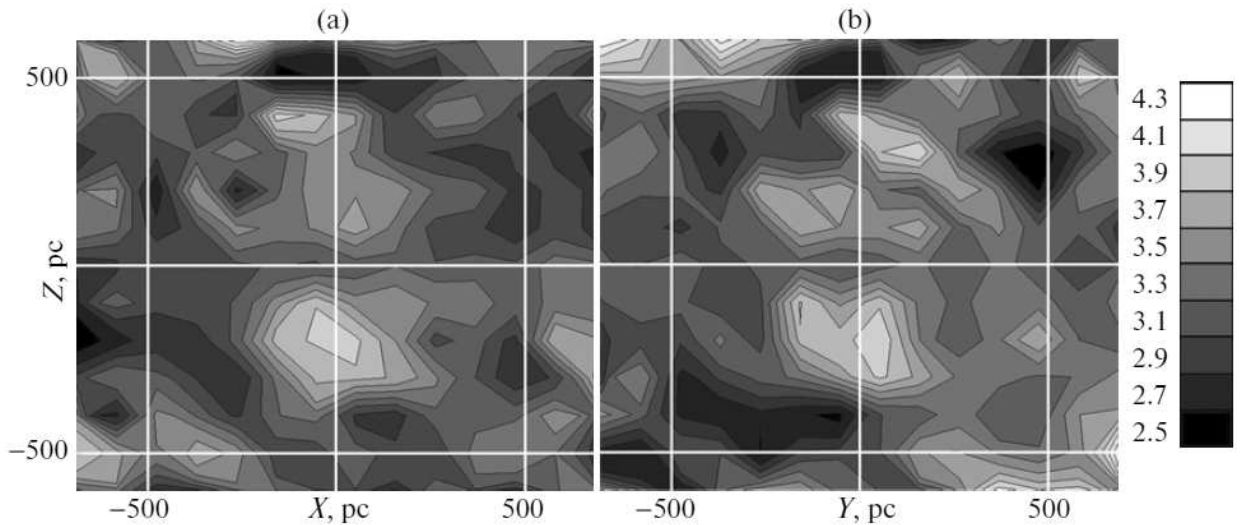}
\caption{Contour maps of $R_V$ as a function of (a) $X$ and $Z$ for the layer
$-150<Y<+150$ pc and (b) $Y$ and $Z$ for the layer $-150<X<+150$ pc based on the data of
Gontcharov (2012a). A color scale of the values of $R_V$ is shown on the right.
The step size for the white lines of the coordinate grid is 500 pc.
The Sun is at the center of the plots.}
\label{fig04}
\end{figure}

Gontcharov (2012a) has compiled a 3D map of the variation in the extinction law (the coefficient $R_V$) within
a radius of 700 pc from the Sun with an accuracy of$\sigma(R_V)=0.2$ and a resolution of 100 pc. As an example, Fig. 4 
shows contour plots of $R_V$ as a function of the coordinates (a) $X$ and $Z$ for the layer $-150<Y<+150$ pc and
(b) $Y$ and $Z$ for the layer $-150<X<+150$ pc. A color scale for the values of $R_V$ is shown on the right. The Sun is
at the center of the plots. Two regions of extremely high RV can be seen to the north and south of the Sun. The
northern region has a ``defect'' in the second quadrant: a zone with lower values of $R_V$ that was noted previously by
Fitzpatrick and Massa (2007). It can also be seen that the spatial variations in $R_V$ are substantially radial relative to
the center of the Gould belt. For example, the reduction in $R_V$ at $|X|\approx500$ pc corresponds roughly to the
position
of the edges of the Gould belt. In addition, in the $XZ$ plane the central layer is inclined to the galactic equator by
roughly 19$^{\circ}$, i.e., as is the Gould belt. The slope of this layer suggests that the spherical structure of the variations
in $R_V$ observed within the entire space considered here is not a heliocentric artifact.

Two samples were used to compile this map: one of 11990 OB stars and the other of 30671 branch giants
of class III. In all the spatial cells that were examined, the values of $R_V$ are consistent with these two samples to
within $\Delta R_V<0.2$. Thus, a final 3D map was obtained by averaging the results over the samples. The agreement
between
the variations in $R_V$ for the red giants and OB stars confirms the similarity of $R_V$ for blue and red stars found by
Berdnikov, et al. (1996). It is clear that the dependences of $R_V$ on reddening, extinction, spectral class, and other
characteristics found by various researchers (see Straizys (1977)) may be, to some extent, artifacts arising from
neglected correlations between the characteristics of the stars and the variations in $R_V$. Evidently, selectivity of the
stars with respect to distance in the catalogs that are limited in stellar magnitude plays a decisive role here: stars with
different colors are observed at different average distances so they have different average values of $R_V$.

\begin{figure}
\includegraphics{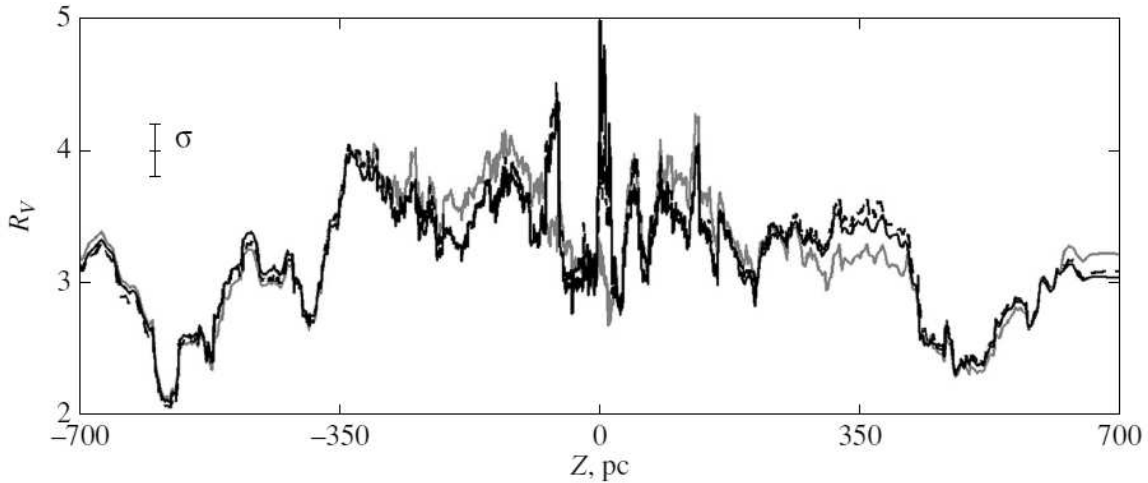}
\caption{Variations in $1.12E(V_T-K_s)/E(B_T-V_T)$ (grey curve),
$1.07E(V_T-W3)/E(B_T-V_T)$ (black dotted curve), and $1.02E(V_T-W4)/E(B_T-V_T)$
(black smooth curve) as functions of $|Z|$ based on the data of Gontcharov
(2016). The error bar indicates the accuracy.}
\label{fig05}
\end{figure}

In Fig. 5 the grey curve, black dashed curve, and black smooth curve show the variations of the approximations
for $R_V$:
$R_V=1.12E(V_T-K_s)/E(B_T-V_T)$,
$R_V=1.07E(V_T-W3)/E(B_T-V_T)$ and
$R_V=1.02E(V_T-W4)/E(B_T-V_T)$.
These were
derived by Gontcharov (2016) on the basis of photometry of 1355 branch giants of class KIII from the Tycho-2,
2MASS, and WISE catalogs in a three-dimensional cylinder along the $Z$ axis with a radius of 150 pc around the
Sun. In these formulas, the coefficient 1.12 was derived from theory, and the coefficients 1.07 and 1.02 have been
selected so that the three values yield the same average value ($\overline{R_V=3.38}$) within the zone $|Z|<700$
pc considered here. The coefficients 1.07 and 1.02 are consistent with their analogs that follow from the WD2001
extinction law, 1.09 and 1.02.
The corresponding extinction $A_{W3}=0.074A_V$ and $A_{W4}=0.027A_V$ (for $\lambda=11$ and 22 $\mu$m)
is caused by silicates (Li (2005), Bochkarev (2009)).

The curves in Fig. 5 diverge significantly only for$-200<Z<-50$, $-2<Z<14$, $100<Z<125$
and $280<Z<400$ pc. In these regions, the spatial mass density of large grains clearly differs from the average. The spike
in the black curves for $-2<Z<14$ pc seems to reflect an elevated abundance of coarse grains in the nearest galactic
surroundings of the solar system, predominantly to the north of the galactic plane. This result is consistent with the
data of Kr\"uger, et al. (2015), and Strub, et al. (2015), who used data from the dust detector in the Ulysses spacecraft
to discover an elevated concentration of large grains in the flow of interstellar dust entering the solar system precisely
from the northern galactic hemisphere together with a flux of neutral hydrogen and helium. (A set of the youngest
OB stars belonging to the Gould belt are moving in the same direction and with roughly the same velocity, about
20 km/s, relative to the Sun (Gontcharov (2012c)). Furthermore, according to Gontcharov (2012b), the galactic
coordinates of the point at which this flux enters the solar system (about $l=3^{\circ}$, $b=21^{\circ}$) and the opposite
point at which it emerges (about $l=183^{\circ}$, $b=-21^{\circ}$) correspond roughly to the regions of maximum
extinction in the Gould belt ($l=15^{\circ}$, $b=19^{\circ}$ and $l=195^{\circ}$, $b=-19^{\circ}$).
Thus, studies of dust grains in the solar system and photometry of stars in its
surroundings have provided a consistent confirmation of the relationship between the solar system and the Gould
belt as a container of coarse dust, gas, and young stars discovered by Taylor, et al. (1996).

Zasowski, et al. (2009), and Gao, et al. (2009), were the first to reliably detect large-scale (over many kpc)
systematic spatial variations in the extinction law within the diffuse matter of the Galaxy. These studies used a very
promising version of a method that was only later fully formulated by Majewski, et al. (2011), for extrapolating the
extinction law employing photometry only in the near and mid IR. In the inner (relative to the Sun) part of the disk,
Zasowski, et al. (2009), and Gao, et al. (2009), found that with distance from the center of the Galaxy, the ratio of
the extintion in the IR to the extinction in the visible decreases (i.e., less IR extinction and a steeper wavelength
dependence in the extinction curve correspond to a less dense medium). If the extinction law is mainly determined
by the size of the dust particles, then finer dust will correspond to lower extinction in the IR. But the same authors
have found a flatter (i.e., the ratio of the extinction in the IR to the absorption in the visible is huge) extinction law
in interstellar space than in the arms. This means that for the most rarefied medium, the IR extinction increases
as the density of the medium falls. Berlind, et al. (1997), found the same dependence on determining the extinction
law in the visible and near IR (0.3-2 $\mu$m) based on photometry of the galaxy IC2163, which is partially eclipsed
by the galaxy NGC2207. They discovered a less flat law in the spiral arms with $A_V\approx1^m$ and a flatter law in the
space between the arms with $A_V\approx0.5^m$.

\begin{figure}
\includegraphics{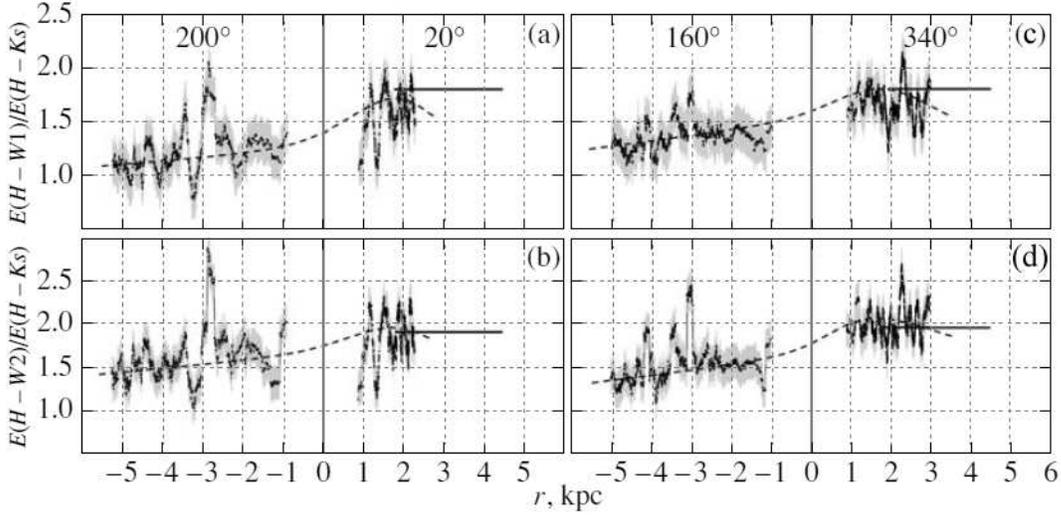}
\caption{$E(V_T-W1)/E(H-K_s)$ and $E(V_T-W2)/E(H-K_s)$ as functions of the distance to
the Sun for longitudes of 200$^{\circ}$ (negative $r$) and 20$^{\circ}$ (positive $r$),
((a)-(b)), 160$^{\circ}$ (negative $r$) and 340$^{\circ}$ (positive $r$) ((c)-(d)):
data from Gontcharov (2013a) shown as black curves with gray error bands.
The horizontal straight lines indicate the analogous results from Zasowski, et al. (2009).
The dashed curves show the approximate systematic variation in the characteristics.}
\label{fig06}
\end{figure}

Gontcharov (2013a) and Schultheis, et al. (2015), have confirmed the results of Zasowski, et al. (2009), and
Gao, et al. (2009), for the inner (relative to the Sun) part of the galactic disk using other photometric data, but found
the opposite behavior in its inner part. Here, with distance from the galactic center over many kpc, the ratio of the
extinction in the IR to that in the visible increases. Thus, newly discovered spatial variations in the extinction law
in the galactic disk are primarily nonmonotonic variations in the dependence on galactocentric distance. Thus, at
some galactocentric distance, slightly closer to the center of the Galaxy than the Sun, the absorption in the IR is
minimal, the wavelength dependence of the extinction is steepest, and the size of the dust particles is minimal. This
can be seen in Fig. 6, which shows the variations $E(H-W1)/E(H-K_s)$ and $E(H-W2)/E(H-K_s)$ in the characteristics of
the extinction law as function of the distance to the Sun for galactic longitudes of
$200^{\circ}$, $20^{\circ}$, $160^{\circ}$ and $340^{\circ}$, i.e.,
near the directions to the center and anticenter of the Galaxy: the curves show the results of Gontcharov (2013a) and
the horizontal lines, the results of Zasowski, et al. (2009), averaged by the authors for the corresponding longitude.

Hutton, et al. (2015), have found the same behavior in the galaxy M82 ($R_V$ initially decreases with distance
from the galactic center and then increases for many kpc).

Gontcharov (2012a, 2013b, 2016) has analyzed the extinction law in spatial cones of height 25 kpc extending
from the Sun to the galactic poles. This made use of photometric data on 9769 clump giants (in a radius of 8$^{\circ}$
around the poles) and 1221 branch giants (in a radius of 20$^{\circ}$ around the poles) from the Tycho-2, 2MASS, and WISE
catalogs. It turned out that outside the galactic plane, at distances $100<|Z|<25000$ pc from it ($Z$ is the coordinate
in the direction of the galactic north pole), everywhere, except thin layers at $Z\approx-600$ ïê and $Z\approx500$ pc,
the extinction law is flatter than near the galactic plane. Figure 7 shows the variations in $E(H-W1)/E(H-K_s)$ and
$E(H-W2)/E(H-K_s)$ with $|Z|$ in the direction of the north (left) and south (right) galactic poles based on data for branch
giants (grey curves) and clump giants (black curves). There is good agreement between the corresponding black and
grey curves within the common range of distances. On the average, for the two poles $E(H-W1)/E(H-K_s)\approx0.8$,
$E(H-W2)/E(H-K_s)\approx0.85$ with spatial variations in these quantities of $\pm0.2$.
For comparison, the same characteristics
obtained by Zasowski, et al. (2009), and Gontcharov (2013a) near the galactic plane, averaged over all longitudes
far from the directions of the galactic center and anticenter, are $E(H-W1)/E(H-K_s)\approx1.7$,
$E(H-W2)/E(H-K_s)\approx2.0$ with spatial variations of $\pm0.4$.
These values are indicated in Fig. 7 by circles and the range of the variations, by the error bars.
The WD2001 model for $R_V=3.1$ gives $E(H-W1)/E(H-K_s)=1.92$, $E(H-W2)/E(H-K_s)=2.37$ for these
characteristics; within the limits of variation, these agree with the observed dependences near the galactic plane, but
not with the ones far from it.
In other words, near the galactic plane $E(H-W1)/E(H-K_s)>1.1$ and $E(H-W2)/E(H-K_s)>1.1$ (i.e., the
dust is relatively fine) almost everywhere, but far from it $E(H-W1)/E(H-K_s)<1.1$ and $E(H-W2)/E(H-K_s)<1.1$
(the dust is coarser) almost everywhere. Thus, far from the galactic plane out to the galactic halo, the IR extinction law does
not look like the extinction law near the plane.

\begin{figure}
\includegraphics{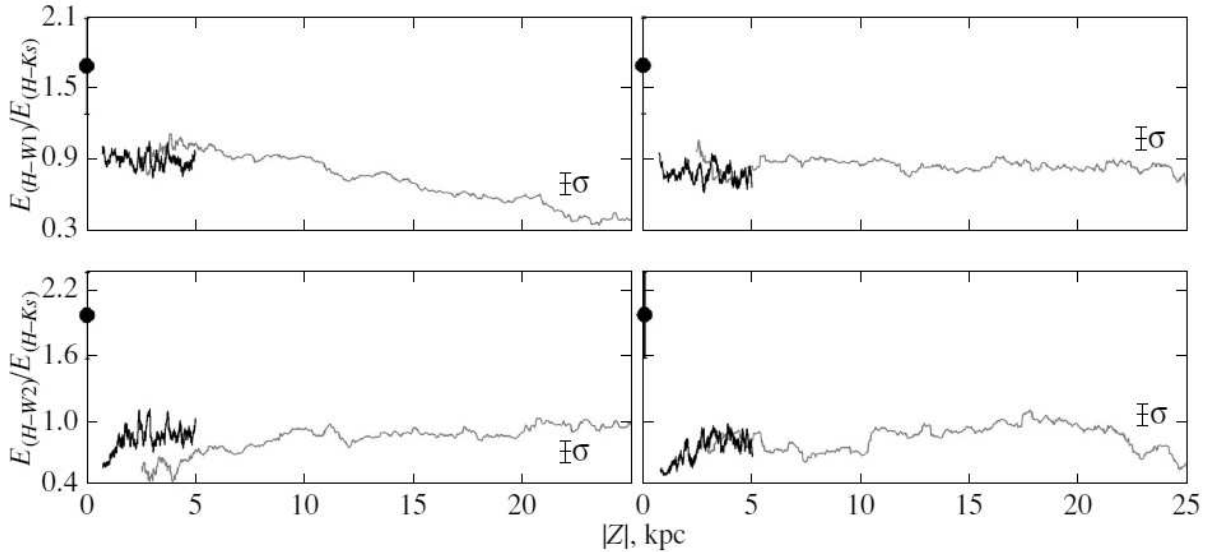}
\caption{Variations in $E(V_T-W1)/E(H-K_s)$ and $E(V_T-W2)/E(H-K_s)$ in the direction of
the northern (left) and southern (right) galactic poles according to the data of
Gontcharov (2013b, 2016) for giants in the branch (grey curves) and clump giants (black curves).
The circles show the characteristics obtained by Zasowski, et al. (2009), and Gontcharov (2013a)
near the galactic plane, averaged over all longitudes far from the directions to the center and
anticenter of the Galaxy. Their spread of $\pm0.4$ is indicated by the error bars.}
\label{fig07}
\end{figure}

The extinction law in the IR has been studied by many authors. Wang, et al. (2014, 2015) have
reviewed the observations and the corresponding models, respectively. Their comparison assumes the existence of coarse
dust particles in the medium. Apparently, these create the emission at wavelengths up to 2 $\mu$m found with the
COBE/DIRBE, COBE/FIRAS, and Planck telescopes. Besides WD2001, one of the models that best fits the observations
is that of Zubko, et al. (2004), with complex dust particles including amorphous and crystalline silicates,
graphite, amorphous carbon (soot), and a shell of organic materials (including polycyclic aromatic hydrocarbons
(PAH)) and water ice, as well as cavities (voids).

Large-scale spatial variations in the extinction law are most noticeable in the IR, since it is possible to look
at the most dusty corners of the Galaxy in the IR. For examples, large variations in the extinction law and in the
properties of the medium have been observed in the galactic bulge, i.e., in the central region with a radius of 2 kpc.
An unusual extinction law has been found before by Popowski (2000) and Sumi (2004). But only as a result of many
studies, e.g., that of Nataf, et al. (2013), has a complete picture developed. With increasing distance from the galactic
center inside the bulge, the average dust particle size initially increases and then decreases, and at the periphery of
the bulge, it transforms into the above-mentioned variations found for the disk by Zasowski, et al. (2009).

Up to now, almost all observations of the extinction law in the IR apply to dense clouds near the galactic plane
or to the galactic center. Here results that agree to within the limits of error have been found by Lutz (1999),
Indebetouw, et al. (2005), Jiang, et al. (2006), Flaherty, et al. (2007), Nishiyama, et al. (2009), Fritz, et al. (2011),
Chen, et al. (2013), and Gao, et al. (2009). The averaged extinction law according to these papers is indicated as
a function of $1/\lambda$ by the circles in Fig. 8 for the $W4$, $W3$, Spitzer/IRAC 8 $\mu$m, Spitzer/IRAC 5.8 $\mu$m,
$W2$ (or Spitzer/IRAC 4.5 $\mu$m) , $W1$ (or Spitzer/IRAC 3.6 $\mu$m), $K_s$, and $H$ bands. As a comparison, the
WD2001 laws for $R_V=3.1$ and 5.5 are plotted, respectively, as the lower and upper dashed curves.
The observations fit the WD2001 law well for $R_V=5.5$. The same IR extinction law, but with a peak at 4.5 $\mu$m,
has been obtained by Wang, et al. (2013), in the Coal Sack nebula and by Gao, et al. (2013), in the dense medium of the
Large Magellanic cloud. These
extinction peaks at 4.5 $\mu$m ($W2$) are indicated in Fig. 8 by the grey and black triangles.
In a review of the properties
of dust and extinction in the Local group of galaxies, Li, et al. (2016), pointed out that all the observations in the
$2-6$ $\mu$m range yield a flat extinction law for dense, as well as diffuse, media.

All of the extinction laws examined here are relative. Thus, the zero point, i.e., the extinction in some band
taken as true, is unique to each study. In order to compare the results, since most of them in the IR agree with the
WD2001 extinction law for $R_V=5.5$, this is taken as the zero point here. At present, the uncertainty in the zero point
of the extinction law is a major problem in all studies. It can be solved in the future only if uniform, high-precision
spectrophotometry is used over a very broad range from the ultraviolet to the far IR.

With the same zero point, the results of Gontcharov (2013b, 2016) far from the galactic plane correspond to
$A_H=0.18A_V$, $A_{K_s}=0.17A_V$, $A_{W1}=0.16A_V$, $A_{W2}=0.16A_V$, $A_{W3}=0.074A_V$, $A_{W4}=0.027A_V$.
The accuracy
of these estimates is $0.03A_V$. These values are indicated in Fig. 8 by the solid diamonds and the solid broken line.
Thus, we see a very flat extinction law in the bands from $H$ through $W2$, i.e., over the range from 1.4 to 5.4 $\mu$m.

It can be seen in Fig. 8 that the extinction in the galactic halo found by Gontcharov (2013b, 2016) in the
$W1$ and $W2$ bands exceeds the other results mentioned here by more than the declared errors. However the results
of Gontcharov (2013b, 2016) were obtained over a much greater distance from the galactic plane, for which there
are almost no other studies for comparison.

Gorbikov and Brosch (2010) have reviewed ``grey'' extinction in various regions of the Galaxy and outside
it. Only one their result applies to the space far from the galactic plane: in the direction of three high-latitude clouds at
distances of $0.5-1$ kpc from the galactic plane they found ``grey'' extinction by $0.2^m-0.4^m$ using SDSS data, a value
in agreement with the results of Gontcharov (2013b, 2016).

Since then, only Davenport, et al. (2014), have analyzed the IR extinction law in the diffuse medium far from
the galactic plane (at $Z$ on the order of 1 kpc) and compared it with the extinction law near the plane. Here 10--band
photometry of more than a million stars from SDSS, 2MASS, and WISE was used. Their results for $|b|>50^{\circ}$
are indicated in Fig. 8 by the open diamonds and for $|b|<25^{\circ}$, by the open squares. The closeness of the circles
and open diamonds in Fig. 8 suggests a similarity in the interstellar medium in the dense clouds near the galactic
plane and in the rarefied medium far from it.

\begin{figure}
\includegraphics{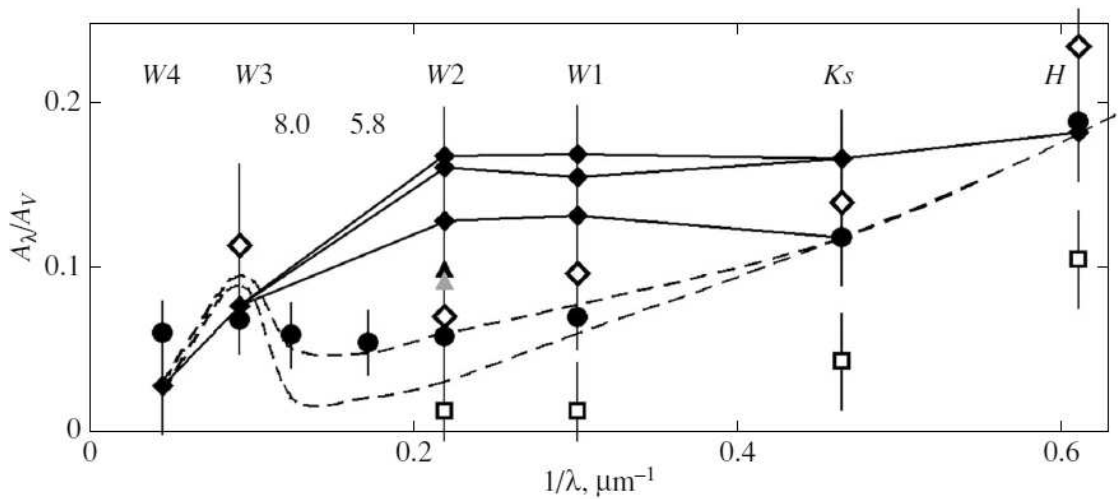}
\caption{The ratio $A_{\lambda}/A_V$ as a function of $1/\lambda$. The WD2001
extinction law for $R_V=3.1$ and $R_V=5.5$ is shown as the lower and upper grey dashed
curves, respectively. The averaged results given in the texts are indicated by the circles.
The grey and black triangles indicate the extinction peaks at 4.5 $\mu$m found by
Wang, et al. (2013), and Gao, et al. (2013). The data of Davenport, et al. (2014), for
$|b|>50^{\circ}$ are indicated by the open diamonds and those for $|b|<25^{\circ}$ by the
open squares.
The extinction law found by Gontcharov (2013b, 2016) for the galactic halo ($5<|Z|<25$ kpc) is
indicated by the diamonds and the black curve. The spectral bands are indicated at the top.}
\label{fig08}
\end{figure}

We note that for latitudes $25^{\circ}<b<50^{\circ}$ Davenport, et al. (2014) have found an especially strong increase in
extinction (relative to $A_V$) in the $H$ and $K_s$ bands. At latitudes $50^{\circ}<b<90^{\circ}$, the relative extinction
in the $H$ and $K_s$ bands decreases, but the extinction increases strongly in $W1$ and $W2$. The same has been shown
by Gontcharov (2013b, 2016). This shows up as a gradual increase in the average size of the dust particles with distance
from the galactic plane. The data of Davenport, et al. (2014), are material for further more detailed analysis.

Schlafly, et al. (2016), have pointed out that in many studies that claim to examine large-scale variations in
the extinction law at high latitudes, a single narrow range of wavelengths and/or a small number of OB stars have
been used, which naturally lie within a narrow layer near the galactic plane; thus, only a few properties of specific
stars and media in a negligibly small part of the Galaxy have been analyzed. One example is a much cited (more
than 70 times) paper of Fitzpatrick and Massa (2009) that analyzed only 14 OB stars within $|Z|<100$ pc. As a source
of confusion, it is worth mentioning a study by Larson and Whittet (2005) who studied the extinction law at high
galactic latitudes but within 100 pc of the Sun. Their results obviously apply to an equatorial dust layer with an
ordinary extinction law.

The variations in the extinction law discussed here, especially in the IR, have been interpreted by Wang, et
al. (2013), in terms of three kinds of medium in the Galaxy: dense clouds of the disk and bulge with coarse dust and
large IR extinction, the diffuse medium of the spiral arms with fine dust and low IR extinction, and the most rarefied
medium between the arms and far from the galactic plane with an extinction law similar to that in dense
clouds. Wang, et al. (2013), emphasize the similarity of the first and third types of media: the temperature is very
low, the gas is molecular, and there are few charged particles. It is evident that in dense clouds, amalgamation of
dust particles predominates over their breakup, in the spiral arms breakup predominates over amalgamation because
of the high temperature, and in the most rarefied medium there are no processes leading to breakup of dust particles.
As confirmation of this, Miville-Deschenes, et al. (2002), have shown that in a typical fairly rarefied high-latitude
cloud the relative velocities of dust particles with a radius on the order of 0.1 $\mu$m are close to the critical velocity
of about 1 km/s (for lower velocities, dust particles merge upon colliding; at higher velocities, they break up). This
hypothesis regarding the three types of medium in the Galaxy requires further testing.

\section*{Coarse dust outside the galactic disk}

It is, therefore, clear that at the periphery of our and other galaxies and, possibly, in the intergalactic medium,
the fraction of coarse dust is greater than in the disks of galaxies. This dust creates the long-known excess radiation
from some extragalactic objects in the far IR at $\lambda\approx500$ $\mu$m.
This emission was detected in observations with the
Herschel and Planck telescopes (Galliano, et al. (2011), Planck (2011b)). Right after its discovery in the 1990's,
attempts were made to interpret this excess as an elevated spatial mass density of cold (with a temperature of $4-7$
K) dust (Reach, et al. (1995)). But it was considered impossible to have such a high spatial density, as well as coarse
dust particles, far from the galactic plane. Modern multicolor IR photometry shows that this is possible, primarily
because of consolidation of dust particles. With a constant spatial mass density, an increase in the radius of a dust
particle by an order of magnitude (e.g., from 0.1 to 1 $\mu$m) leads to a reduction in the number of particles per unit
volume by 3 orders of magnitude and to a corresponding increase in the average distance between particles. This
greatly increases the transparency of the medium and reduces the interactions of matter with radiation. Coarsening
of the dust particles leads to an increase in their far IR emission. At present, this is the only explanation for the
anticorrelation between the IR radiation and the density of the medium in the Large Magellanic cloud discovered
by Galliano, et al. (2011), based on data from the Herschel telescope.

We note that the above mentioned emission in the mid IR range found by Yahata, et al. (2007), and Wolf
(2014) in a comparison of photometry of galaxies and quasars with the SFD98 map may be a manifestation of
previously neglected medium-sized dust particles. And the unexpected discovery by the Cosmic Infrared Background
ExpeRiment (CIBER) rocket-borne experiment of powerful emission in the near IR from the peripheries of galaxies
and from intergalactic space (Zemcov, et al. (2014)) is evidently a manifestation of neglected fine dust. Thus, it is
possible that up to now the spatial mass density of all dust (and not just the coarse dust) has been underestimated.

The large fraction of coarse dust in the medium is confirmed by observations of the halo caused by this dust
around X-ray sources. Witt, et al. (2001), detected a halo created by coarse interstellar dust particles (radii at least
2 $\mu$m) around the Nova Cygni 1992 X-ray source. Corrales and Paerels (2012) found an X-ray halo around the Cygnus
X-3 source that was probably created by coarse dust particles in the interstellar medium of the Cyg OB2 association
against the background of the source.

According to data from the Planck collaboration (2011c), up to 90.7\% of the entire mass of dust at the
periphery of the galaxy consists of coarse grains. There the increased size and mass of solid particles facilitates a
constant, low temperature (on the order of 10 K according to Planck (2011a)) and similar kinetic energies of the
objects, but also an increased role for Van der Waals forces compared to gravity, the accumulation of ice shells on
the surface of grains, and a high gas absorption capacity of these particles (Sandford and Allamandola (1993)) so that
atomic hydrogen is converted into molecules (Dissly, et al. (1994); Perets, et al. (2005)). The rate of this conversion
on the surface of the particles increases with falling temperature and increasing particle size (Lipshtat and Biham
(2005)). The accumulation of ice shells on refractory grains and the appearance of layered grains also favor increased
extinction in the near IR (Fritz, et al. (2011); Voshchinnikov (2012)).

Greenberg (1974) was the first to point out that the interstellar medium is well described only by models
containing coarse (greater than 0.1 $\mu$m) dust. This has been confirmed in the 21-st century. Voshchinnikov (2012)
has reviewed the various grain size distributions that have been proposed. For example, as we have seen, the WD2001
model corresponds best to observations in the $2-8$ $\mu$m range if it is assumed that $R_V=5.5$. But then it includes a
substantial fraction of carbon gains with radii on the order of 0.5 $\mu$m and, in some versions, up to 7.3 $\mu$m.
Incidentally, the authors of WD2001 acknowledge that the model cannot explain (with any parameters) the very large
fraction of coarse (radii on the order of 0.5 $\mu$m) interstellar dust particles detected by the Ulysses and Galileo
spacecraft in the solar system according to the WD2001 data. Frisch, et al. (1999), pointed out the need for further
observations of coarse dust within and outside the solar system. These observations have been made and are discussed
in the next section.

According to modern ideas, dust is formed in the shells of red giants, supergiants, novae and supernovae and
is transported from them into the interstellar medium (Bochkarev (2009)). But for a long time, observational data
on the dust mass produced this way differed by several orders of magnitude from theoretical estimates (Wesson, et
al. (2015)). In recent years, because of microwave observations of these stars, these estimates have been substantially
revised.

A large mass of dust (on the order of $0.5M_{\odot}$) was detected for the first time during the outburst of the
SN1987A supernova by Matsuura, et al. (2011), based on far IR observations with the Herschel space observatory.
25 years after the outburst, Wesson, et al. (2015), discovered dust with an overall mass of $0.6-0.8M_{\odot}$ around
SN1987A
based on Herschel observations. Getting agreement with the observed spectral energy distribution will require a
substantial fraction of dust grains with radii greater than 2 $\mu$m. They concluded that a substantial growth of grains
occurred during a certain period. Right after them, and also using observations from Herschel, Matsuura, et al. (2015),
explained the observed SED of the supernova SN1987A in terms of the formation of silicate
and amorphous carbon dust particles with masses of 0.5 and $0.3M_{\odot}$, respectively (a combined dust mass of about
$0.8M_{\odot}$). Indebetouw, et al. (2014), have analyzed the cold dust shell (the remnant of the core of the exploded star)
observed around SN1987A with the ALMA (Atacama Large Millimeter/Submillimeter Array) telescope, which has
better resolution than the Herschel, and found that all the carbon produced by the star had very efficiently condensed
into dust. They concluded that, over the first 25 years, this supernova produced dust with a total mass on the order
of $0.2M_{\odot}$ and that the bulk of this dust (primarily coarse grains) should apparently move on into the interstellar
medium. Relying on the results of Dwek, et al., (2007), they concluded that, in this case, supernovae with a collapsing
core are the dominant sources of dust in the medium of galaxies with any red shifts. Gall, et al. (2014), have
analyzed the properties of the dense medium surrounding the supernova SN 2010jl a few days after an explosion.
They concluded that dust particles with radii from 0.001 to 4.2 $\mu$m appeared in this medium, with most larger than
1 $\mu$m (80\% of the mass of the dust is contained in grains with radii greater than 0.1 $\mu$m). The size
distribution of these dust particles follow a power-law distribution with an exponent of 3.6. It is shown in Fig. 9 and is
discussed below.
According to the data of Gall, et al. (2014), the dust from the supernovae SN 1995N, SN 1998S, SN 2005ip and
SN2006jd had roughly the same characteristics. Scicluna, et al. (2015), also found coarse dust using the SPHERE
equipment at the VLT telescope in an analysis of gas-dust shell shed by the supergiant VY Canis Majoris. The mass
loss is $0.0001M_{\odot}$ per year. The average radius of the grains is 0.5 $\mu$m. A high radiation pressure must drive
these large grains into the interstellar medium without loss and destruction.

\begin{figure}
\includegraphics{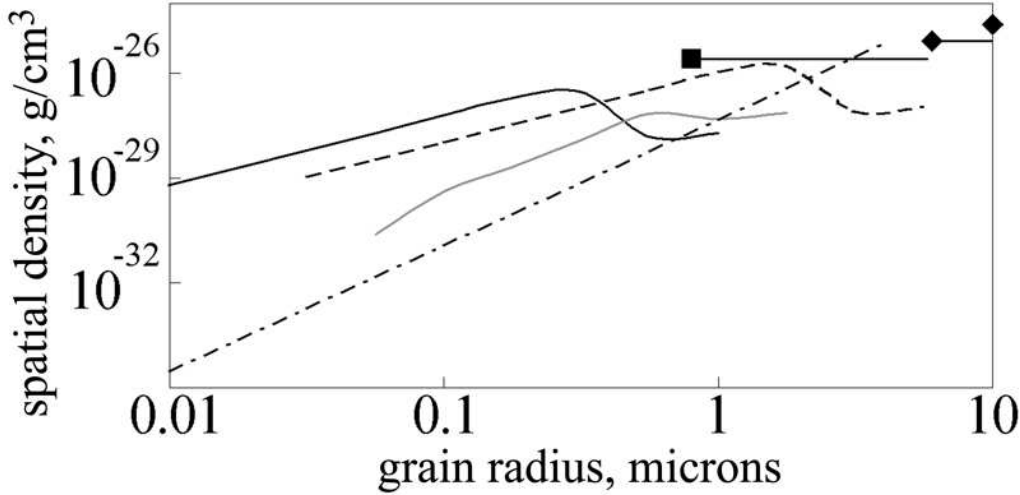}
\caption{The distribution of the spatial density of dust as a function of dust particle radius in
the shell of a supernova (dot-dashed curve), in interstellar medium with carbon and silicate dust
(smooth black curve), in interstellar medium with ice dust (different normalization) (black
dashed curve), and inside the solar system according to data from the Ulysses spacecraft
(grey smooth curve), according to data from the New Horizons spacecraft (square with an
indication of uncertainty),
and according to data from the Pioneer spacecraft (diamonds with an
indication of uncertainty).}
\label{fig09}
\end{figure}

Li, et al. (2011), have estimated an average frequency of supernovae with collapsing cores in the Galaxy as
2.3 per century. If a supernova produces dust with a mass of $0.5M_{\odot}$ on the average, then the medium is enriched
in dust at a rate of $0.011M_{\odot}$ per year. (Similar calculations for the Large Magellanic cloud have been done by
Matsuura, et al. (2011).) On the other hand, for a star formation rate on the order of $1.6M_{\odot}$ per year in the Galaxy
(Licquia and Newman, 2015) and a ratio of the masses of gas and dust of 100, we can see that star formation consumes
on the order of $0.016M_{\odot}$ per year from the medium. Thus, the production of dust (including coarse particles) by
supernovae (even neglecting giants and supergiants) is now sufficient to maintain a constant average spatial density
of the interstellar medium, despite steady consumption of the medium in the formation of stars and planets.

\section*{Dust inside and outside the solar system}

In recent years data have become available which make it possible to compare the size distribution and total
mass of dust (1) in the places where dust is formed, (2) in the interstellar medium, and (3) at the edge and (4) inside
the Solar system.

Extinction/emission can be used to estimate the size of dust grains, and impact on spacecraft has made it
possible to estimate the mass the dust grains. To compare these results, it is necessary to know the average physical
density of dust grains. At present, the only small solid objects for which the density is reliably known, directly
observable, and formed beyond the orbit of Jupiter (i.e., the closest to interstellar dust particles in terms
of their ambient conditions) are the regular satellites of Saturn (Thomas, 2010) and Uranus (Jacobson, et al. (1992))
with densities of about 1 g/cm$^3$. There are other reasons (Dwek, et al. (2007); Bochkarev (2009), p. 296) for assuming
here and afterward that this is the average physical density for interstellar dust particles.

Let us use the value in Eq. (1) to normalize the size distribution of dust particles produced by the supernova
SN 2010jl according to Gall, et al. (2014). This can be regarded as \emph{the size distribution of dust particles in the
locations where they are produced} and is plotted as the dot-dashed curve of Fig. 9.

The spatial mass density of dust as a function of particle radius according to the WD2001 model for $R_V=5.5$
is plotted as the smooth black curve in Fig. 9. Casuso and Beckman (2010) have shown that this model distribution
agrees in order of magnitude with the observed distributions for the galaxy and the Magellanic clouds. It can be
taken as \emph{the size distribution of dust grains in the interstellar medium}.
A large fraction of coarse dust has been found
in supernovae and in the solar system, so a variant of the WD2001 model with a maximum fraction of coarse dust,
i.e., with maximum $R_V$, has been chosen.

Impacts of interstellar dust particles detected on the Ulysses and Galileo spacecraft (Frisch, et al. (1999);
Kr\"uger, et al. (2001)) show that dust particles with masses from $10^{-12}$ to $2\cdot10^{-12}$ g, i.e., with radii on
the order of 0.7 $\mu$m, form the bulk of the dust mass. Kr\"uger, et al. (2001), have obtained an estimate for the spatial
density ($2.1\cdot10^{-27}$ g/cm$^3$) and a size distribution of the interstellar dust penetrating into the solar system
(the characteristics of this flux are discussed above). This estimate was derived from 16 years of data acquired by a dust
detector on the Ulysses spacecraft far from the ecliptic at a heliocentric distance of about 5 AU and is roughly an order of
magnitude lower than the value in Eq. (1) for interstellar space. This distribution is plotted as the smooth grey curve
in Fig. 9 and is \emph{an estimate for dust inside the solar system}. Here its radius has been calculated from a physical
density
of 1 g/cm$^3$. Kr\"uger, et al. (2001), point out that the data from Ulysses are consistent with data from Galileo (which
detects dust mainly at heliocentric distances of about 5 AU), Cassini ($1-9.5$ AU), and Helios ($0.3-1$ AU).

We note that the solar system is full of fluxes of interplanetary dust of local origin; sometimes these
are large (owing to the destruction of asteroids and comets, ejection of matter by cryovolcanoes, etc.). For example,
Bauer, et al. (2008), have shown that 90\% of the mass of dust lost by the Echeclus centaur is contained within particles
of size of the order of 30 $\mu$m. This loss of dust is not the result of a collision from outside (although the loss
mechanism is not clear), otherwise the fraction of coarse dust would be even greater. But in the data from these
spacecraft, the interstellar and interplanetary dust particles are reliably separated.

According to Popp, et al. (2010), the dust sensor on board the New Horizons spacecraft at heliocentric
distances of $2.6-15.5$ AU indicated a spatial density of coarse interstellar dust of the same order as the Ulysses, Galileo,
Pioneer 10, Pioneer 11, Voyager 1, and Voyager 2 spacecrafts at comparable heliocentric distances. At heliocentric
distances of $6.8-15.5$ AU, for dust particles with masses of $2\cdot10{-12}-10^{-9}$ g (i.e., radii of $0.8-6.2$ $\mu$m),
the spatial
density according to the New Horizons data averaged $2.6\cdot10^{-26}$ g/cm$^3$. This value is indicated in Fig. 9 by a
square
at a radius of 0.8 $\mu$m and the horizontal line emerging from it to show the error in the radius. This can be regarded
as \emph{an estimate for the dust at the edge of the solar system}. The vertical shift in Fig. 9 between the black square and
the grey curve was noted by Popp, et al. (2010), in the New Horizons data and indicates an increase in the flux of
interstellar dust with increasing distance from the Sun beyond the orbit of Jupiter. For comparison Popp, et al. (2010),
list the spatial density of all the dust (interplanetary and interstellar) beyond the orbit of Jupiter based on data from
the Pioneer spacecraft: $8.6\cdot10^{-26}$ g/cm$^3$ for grains more massive than  $8.3\cdot10^{-10}$ g, i.e.,
with radii greater than 6 $\mu$m,
and $2.5\cdot10^{-25}$ g/cm$^3$ for grains more massive than $6\cdot10^{-9}$ g, i.e., with radii greater than 11.2 $\mu$m.
These values are
indicated by the two diamonds in Fig. 9 (the horizontal line to the right of the leftmost diamond shows the error
in the radius). The New Horizons spacecraft did not detect any grains that were so large. The reason, as laboratory
tests by Popp, et al. (2010) show, is the technical limitations of the detector in the New Horizons system. Kr\"uger,
et al. (2015), pointed out the technical limits of the dust detector in the Ulysses spacecraft, which kept it from
detecting collisions with dust grains with masses greater than $3\cdot10^{-11}$ g, i.e., radii greater than 2 $\mu$m.
It is clear that
the bulk of the dust mass in the solar system is contained in grains with radii greater than 0.5 $\mu$m. The degree to
which this result is unexpected can be seen from the fact that the designers of almost all the dust detectors for
spacecraft systems did not expect to detect such coarse dust particles.

The black and grey curves in Fig. 9 are very similar to the analogous curve in Fig. 24 of the article WD2001,
where the WD2001 model is compared with earlier data from spacecraft for the solar system by Frisch, et al. (1999).
Thus, at first glance, the conclusion of WD2001 that the mass distribution of dust particles in the interstellar medium
and the solar system differ radically (like the black and grey curves in Fig. 9) seems justified. WD2001, however,
operates with the mass of the dust particles, while the measured quantity is an extinction/emission wavelength. The
calibration of WD2001 begins with an average grain density of about 3 g/cm$^3$ under the assumption that the fine
grains consist of silicates (3.5 g/cm$^3$) and the coarse grains, of graphite (2.24 g/cm$^3$).
For $R_V=5.5$, the maximum of
the WD2001 distribution at $3\cdot10^{-27}$ g/cm$^3$ applies to grains with a mass of $2\cdot10^{-13}$ g, i.e.,
a radius of 0.25 $\mu$m. Carbon
grains of this size have maximal extinction at 1.5 $\mu$m (Bohren and Huffman, 1986; Bochkarev, 2009). But if the grains
consist predominantly of water ice and have an average density of 1 g/cm$^3$, the peak extinction wavelength over the
size distribution of the grains is different. This calculation is shown below.

The attenuation of the radiation from luminous objects in space, which we traditionally refer to as extinction,
actually includes scattering and intrinsic extinction (although separating these is not always important). They are
described by the real $n'$ and imaginary $n''$ parts of the complex refractive index  $n=n'-i\cdot n''$.
$n'$ and $n''$ can be
complicated functions of $\lambda$ that depend on chemical composition and other characteristics of dust particles
(Bochkarev, 2009).
As shown before, the bulk of the mass of interstellar dust clearly is contained in dust particles that cause
extinction in the $0.6-5$ $\mu$m range (in the WD2001 model with $R_V=5.5$, at 1.5 $\mu$m).
Within this range, $n'\approx1$ (Bochkarev, 2009).
But $n''$ varies over wide limits, for example from $n\ll1$ for water ice to $n''\approx1$ for carbon
(Bohren and Huffman, 1986). This leads to uncertainty in particle size calculations as a function of the peak extinction
wavelength. But, as an approximate estimate we may assume that the ratio of wavelength to the dust particle radius
ranges from approximately 1 for ice particles to 6 for carbon particles (Bohren and Huffman, 1986; Bochkarev, 2009).
Then the maximum extinction at 1.5 $\mu$m may be caused by ice particles with a radius of 1.5 $\mu$m, instead of carbon
particles with a radius of 0.25 $\mu$m. For a density of 1 g/cm$^3$, the mass of such a dust particle is $1.4\cdot10^{-11}$ g.
The entire distribution is shifted in a similar way: it is shown as the dashed curve of Fig. 9. Here the spatial mass density
of the dust is estimated using the normalization of Eq. (1). This shifts the dashed curve along the ordinate by a slight
amount relative to the smooth black curve.

Figure 9 shows that the new distribution (the dashed curve) is much more easily compared with the other
results than is the original (smooth black) curve.
\begin{enumerate}
\item On comparing the dust in the neighborhood of supernovae (dot-dashed curve) and in the interstellar
medium (dashed curve), we can see that the initially fairly coarse dust (on the average) has clearly been broken up
with time.
\item On comparing the dust in the middle (dashed curve) and at the edge of the solar system (the square and
diamonds), there is no discrepancy: the maxima of the distributions agree in order of magnitude. Thus, it is clear
that interstellar dust penetrates into the outer regions of the solar system almost without destruction.
\item On comparing the dust in the interstellar medium (dashed curve) and inside the solar system (grey curve),
we see that a small fraction of the dust penetrates (the barrier mechanisms have been examined by Kr\"uger, et al (2015)),
although the size distribution of the grains is retained to some extent.
\item If typical interstellar dust grains are coated with an ice shell, which forms the bulk of their mass, and if
typical dust has been observed thus far in supernovae and in the solar system, then it is possible that dust particles
with radii greater than 0.5 $\mu$m are produced in and predominate in the medium (possibly except for a thin equatorial
layer in the spiral arms) and (predominantly, in terms of mass) enter the solar system.
\end{enumerate}

\section*{Conclusions}

According to current ideas, the interstellar medium contains roughly half the mass of the matter in the
circumsolar part of the Galaxy. Thus, studies of this medium are very important.

Ideas regarding the distribution of the dust in the layer near the galactic plane ($|Z|<100$ pc) and the related
estimates of extinction in this layer have hardly changed over the last 170 years: from $1^m$ per kpc according to
Struve (1847) to $1.2^m$ per kpc according to Gontcharov (2012b) based on Eq. (12). Thus, the major discoveries relating to
extinction and the dusty medium in the 21st century are associated with studies of regions with $|Z|>100$ pc.
In particular, it has been possible to characterize the extinction numerically in the dust layer in the Gould belt, as well
as in the equatorial dust layer. A comparison of Eqs. (12) and (13) shows that for some longitudes and latitudes,
the extinction in the Gould belt is even greater than in the equatorial layer. Taking this into account in, for example,
the model of Gontcharov (2009) and the sum of Eqs. (12) and (13) is especially important for extragalactic objects.
Then it is sufficient to know the coordinates of an object, and for objects at middle and high latitudes lying well
beyond 600 pc, it is not necessary to know the distance.

The most popular source for dealing with extinction up to now, SFD98, has systematic errors, the most
important of which can be corrected using Eq. (4).

One of the major problems in modern astronomy for the study of dusty media and extinction is improving
the accuracy of the extinction law, especially far from the galactic plane and in the infrared (where the medium is
more transparent and extinction is caused by dust grains that appear to form the bulk of the dust). The extinction
law is now more often determined by solving the system of Eqs. (15) or its analogs, together with a determination
of the key characteristics of each star based on multicolor photometry or spectrophotometry of millions of stars. This
approach provides an order of magnitude more information about a star than spectral classification.

Today, research on dust and the extinction/radiation produced by it covers almost the entire Galaxy and has
revealed substantial variations in the characteristics of the dusty medium and, especially, in the size distribution of
dust grains. Many studies have revealed an unexpectedly large fraction of the coarse dust produced by supernovae
and supergiants, in some regions of the bulge, in the inter-arm space, at the periphery of the disk, in the galactic halo,
within a radius of 14 pc from the sun, and inside the solar system. This requires further testing, especially in
relation with the uncertainty of the cosmological parameters.

I thank V. P. Grinin and N. G. Bochkarev for valuable comments. Much use has been made in this review
of data from the Hipparcos-Tycho, Two Micron All Sky Survey (2MASS) and Wide-field Infrared Survey Explorer
(WISE) programs, as well as the resources of the Strasbourg center for astronomical data (Centre de Donn\'ees
Astronomiques de Strasbourg). This work was supported by program P-7 of the Presidium of the Russian Academy
of Sciences and the subprogram ``Transition and explosive processes in astrophysics.''


\begin{thebibliography}{199}

\bibitem{sdss} Abazajian K. N., Adelman-McCarthy J. K., Agueros M. A., et al., Astrophys. J. Supp. Ser. 182, 543 (2009).
\bibitem{arce} Arce H. G. and Goodman A. A., Astrophys. J. 512, L135, (1999).
\bibitem{arenou} Arenou F., Grenon M., and Gomez A., Astron. Astrophys. 258, 104, (1992).
\bibitem{bauer} Bauer J. M., Choi Y.-J., Weissman P. R., et al., Publ. Astron. Soc. Pacif. 120, 393 (2008).
\bibitem{berdnikov} Berdnikov L. N., Vozyakova O. V., and Dambis A. K., Pis'ma v Astron. zh. 22, 372 (1996).
\bibitem{berlind} Berlind A. A., Quillen A. C., Pogge R. W., et al., Astron. J. 114, 107 (1997).
\bibitem{berry} Berry M., Ivezic Z., Sesar B., et al., Astrophys. J. 757, 166 (2012).
\bibitem{binney} Binney J. and Tremaine S., Galactic Dynamics (Princeton University Press, 2008).
\bibitem{bobylev} Bobylev V. V., Astrofizika 57, 625 (2014), [Astrophysics 57, 583 (2014)].
\bibitem{bsitnik} Bochkarev N. G. and Sitnik T. G., Astron. Astrophys. Suppl. Ser. 108, 237 (1985).
\bibitem{bochka} Bochkarev N. G., Foundations of the Physics of the Interstellar Medium [in Russian], Knizhnyi dom LIBROKOM (2009), pp. 291--337.
\bibitem{bohren} Bohren C. F. and. Huffman D. R., Absorption and Scattering of Light by Small Particles [Russian translation], Mir, Moscow (1986), p. 26.
\bibitem{b2012} Bressan A., Marigo P., Girardi L., et al., Mon. Not. Roy. Astron. Soc. 427, 127 (2012), http://stev.oapd.inaf.it/cmd
\bibitem{cambresy} Cambresy L., Jarrett T. H., and Beichman C. A., Astron. Astrophys. 435, 131, (2005).
\bibitem{cardelli} Cardelli J. A., Clayton G. C., and Mathis J. S., Astrophys. J. 345, 245 (1989).
\bibitem{casuso} Casuso E. and Beckman J. E., Astron. J. 139, 1406 (2010).
\bibitem{chen2013} Chen B.-Q., Schultheis M., Jiang B. W., et al., Astron. Astrophys. 550, A42 (2013).
\bibitem{chen2014} Chen B.-Q., Liu X.-W., Yuan H.-B., et al., Mon. Not. Roy. Astron. Soc. 443, 1192 (2014).
\bibitem{corrales} Corrales L. and Paerels F., Astrophys. J. 751, 93 (2012).
\bibitem{bmg2} Czekaj M. A., Robin A. C., Figueras F., et al., Astron. Astrophys. 564, A102 (2014).
\bibitem{dame} Dame T. M., Ungerechts H., Cohen R. S., et al., Astrophys. J. 322, 706 (1987).
\bibitem{davenport} Davenport J. R. A., Ivezic Z., Becker A. C., et al., Mon. Not. Roy. Astron. Soc. 440, 3430 (2014).
\bibitem{dissly} Dissly R. W., Allen M., and Anicich V. G., Astrophys. J. 435, 685 (1994).
\bibitem{draine} Draine B. T., Ann. Rev. Astron. Astrophys. 41, 241 (2003).
\bibitem{dutra2002} Dutra C. M. and Bica E., Astron. Astrophys. 383, 631 (2002).
\bibitem{dutra2003} Dutra C. M., Santiago B. X., Bica E. L. D., et al., Mon. Not. Roy. Astron. Soc. 338, 253, (2003).
\bibitem{dwek} Dwek E., Galliano F., and Jones A. P., Astrophys. J. 662, 927 (2007).
\bibitem{fm2007} Fitzpatrick E. L. and Massa D., Astrophys. J. 663, 320, (2007).
\bibitem{fm2009} Fitzpatrick E. L. and Massa D., Astrophys. J. 699, 1209 (2009).
\bibitem{flaherty} Flaherty K. M., Pipher J. L., Megeath S. T., et al., Astrophys. J. 663, 1069 (2007).
\bibitem{frisch} Frisch P. C., Dorschner J. M., Geiss J., et al., Astrophys. J. 525, 492 (1999).
\bibitem{fritz} Fritz T. K., Gillessen S., Dodds-Eden K., et al., Astrophys. J. 737, 73 (2011).
\bibitem{gall} Gall C., Hjorth J., Watson D., et al., Nature, 511, 326 (2014).
\bibitem{galliano} Galliano F., Hony S., Bernard J.-P., et al., Astron. Astrophys. 536, A88 (2011).
\bibitem{gao2009} Gao J., Jiang B. W., Li A., Astrophys. J. 707, 89 (2009).
\bibitem{gao2013} Gao J., Jiang B. W., Li A., et al., Astrophys. J. 776, 7 (2013).
\bibitem{gomez2006} Gomez G. C., Astron. J. 132, 2376 (2006).
\bibitem{gould} Gontcharov G. A., Astron. Lett., 35, 780 (2009).
\bibitem{map} Gontcharov G. A., Astron. Lett., 36, 584 (2010).
\bibitem{rv} Gontcharov G. A., Astron. Lett., 38, 12 (2012a).
\bibitem{av} Gontcharov G. A., Astron. Lett., 38, 87 (2012b).
\bibitem{apex} Gontcharov G. A., Astron. Lett., 38, 694 (2012c).
\bibitem{ob} Gontcharov G. A., Astron. Lett., 38, 771 (2012d).
\bibitem{law} Gontcharov G. A., Astron. Lett., 39, 83 (2013a).
\bibitem{law2} Gontcharov G. A., Astron. Lett., 39, 550 (2013b).
\bibitem{law3} Gontcharov G. A., Astron. Lett., 42, 445 (2016).
\bibitem{gorbikov} Gorbikov E. and Brosch N., Mon. Not. Roy. Astron. Soc. 401, 231 (2010).
\bibitem{green} Green G. M., Schlafly E. F., Finkbeiner D. P., et al., Astrophys. J. 783, 114 (2014).
\bibitem{greenberg} Greenberg J. M., Astrophys. J. 189, L81 (1974).
\bibitem{tycho2} H\o g E., Fabricius C., Makarov V. V., et al., Astron. Astrophys. 355, L27 (2000).
\bibitem{hutton} Hutton S., Ferreras I., and Yershov V., Mon. Not. Roy. Astron. Soc. 452, 1412 (2015).
\bibitem{ind2005} Indebetouw R., Mathis J. S., Babler B. L., et al., Astrophys. J. 619, 931 (2005).
\bibitem{ind2014} Indebetouw R., Matsuura M., Dwek E., et al., Astrophys. J. 782, L2 (2014).
\bibitem{jacob} Jacobson R. A., Campbell J. K., Taylor A. H., et al., Astron. J. 103, 2068 (1992).
\bibitem{jiang} Jiang B. W., Gao J., Omont A., et al., Astron. Astrophys. 446, 551 (2006).
\bibitem{johnson} Johnson H. L., Astrophys. J. 141, 923 (1965).
\bibitem{jb} Johnson H. L. and Borgman J., Bulletin of Astronomical Institute of Netherlands, 17, 115 (1963).
\bibitem{jones} Jones D. O., West A. A., and Foster J. B., Astron. J. 142, 44 (2011).
\bibitem{kruger2001} Kr\"uger H., Gr\"un E., Landgraf M., et al., Planetary and Space Sci., 49, 1303 (2001).
\bibitem{kruger2015} Kr\"uger H., Strub P., Gruun E., et al., Astrophys. J. 812, 139 (2015).
\bibitem{kulik} Kulikovskii P. G., Stellar Astronomy [in Russian], Nauka, Moscow (1985).
\bibitem{larson} Larson K. A. and Whittet D. C. B., Astrophys. J. 623, 897 (2005).
\bibitem{li2005} Li A., J. Phys. Conf. Ser. 6, 229 (2005).
\bibitem{li2016} Li A., Wang S., Gao J., et al., in: K. C. Freeman, et al., ed., Lessons from the local group, Springer, New York (2016), p. 85.
\bibitem{li2011} Li W., Chornock R., Leaman J., et al., Mon. Not. Roy. Astron. Soc. 412, 1473 (2011).
\bibitem{licquia} Licquia T. C. and Newman J. A., Astrophys. J. 806, 96 (2015).
\bibitem{lipshtat} Lipshtat A. and Biham O., Mon. Not. Roy. Astron. Soc. 362, 666 (2005).
\bibitem{lutz} Lutz D., The Universe as Seen by ISO, Eds. P. Cox, M. F. Kessler, 623 (1999).
\bibitem{majewski} Majewski S. R., Zasowski G., and Nidever D. L., Astrophys. J. 739, 25 (2011).
\bibitem{matsuura2011} Matsuura M., Dwek E., Meixner M., et al., Science, 333, 1258 (2011).
\bibitem{matsuura2015} Matsuura M., Dwek E., Barlow M. J., et al., Astrophys. J. 800, 50 (2015).
\bibitem{mckee} McKee C. F., Parravano A., and Hollenbach D. J., Astrophys. J. 814, 13 (2015).
\bibitem{miville} Miville-Deschenes M.-A., Boulanger F., Joncas G., et al., Astron. Astrophys. 381, 209 (2002).
\bibitem{nataf} Nataf D. M., Gould A., Fouque P., et al., Astrophys. J. 769, 88 (2013).
\bibitem{nishiyama} Nishiyama S., Tamura M., Hatano H., et al., Astrophys. J. 696, 1407 (2009).
\bibitem{parenago} Parenago P. P., A Course in Stellar Astronomy [in Russian], GITTL, Moscow (1954).
\bibitem{peek} Peek J. E. G. and Graves G. J., Astrophys. J. 719, 415, (2010).
\bibitem{perets} Perets H. B., Biham O., Manico G., et al., Astrophys. J. 627, 850 (2005).
\bibitem{per} Perryman M., Astronomical Application of Astrometry (Cambridge Univ. Press, Cambridge (2009).
\bibitem{planck16} Planck Collaboration, Ade P. A. R., Aghanim N., et al., Astron. Astrophys. 536, A16 (2011a).
\bibitem{planck17} Planck Collaboration, Ade P. A. R., Aghanim N., et al., Astron. Astrophys. 536, A17 (2011b).
\bibitem{planck24} Planck Collaboration, Ade P. A. R., Aghanim N., et al., Astron. Astrophys. 536, A24 (2011c).
\bibitem{popow} Popowski P., Astrophys. J. 528, L9, (2000).
\bibitem{poppe} Poppe A., James D., Jacobsmeyer B., et al., Geophysical Research Letters, 37, L11101 (2010).
\bibitem{reach} Reach W. T., Dwek E., Fixsen D. J., et al., Astrophys. J. 451, 188 (1995).
\bibitem{reis} Reis W. and Corradi W. J. B., Astron. Astrophys. 486, 471 (2008).
\bibitem{robin} Robin A. C., Reyle C., Derriere S., et al., Astron. Astrophys. 409, 523 (2003).
\bibitem{sandford} Sandford S. A. and Allamandola L. J., Astrophys. J. 409, L65 (1993).
\bibitem{schlafly14a} Schlafly E. F., Green G., Finkbeiner D. P., et al., Astrophys. J. 789, 15 (2014a).
\bibitem{schlafly14b} Schlafly E. F., Green G., Finkbeiner D. P., et al., Astrophys. J. 786, 29 (2014b).
\bibitem{schlafly16} Schlafly E. F., Meisner A. M., Stutz A. M., et al., Astrophys. J. 821, 78 (2016).
\bibitem{sfd} Schlegel D. J., Finkbeiner D. P., and Davis M., Astrophys. J. 500, 525 (1998).
\bibitem{schultheis} Schultheis M., Kordopatis G., Recio-Blanco A., et al., Astron. Astrophys. 577, 77 (2015).
\bibitem{scicluna} Scicluna P., Siebenmorgen R., Wesson R., et al., Astron. Astrophys. 584, L10 (2015).
\bibitem{2mass} Skrutskie M. F., Cutri R. M., Stiening R., et al., Astron. J. 131, 1163 (2006). http://www.ipac.caltech.edu/2mass/releases/allsky/index.html
\bibitem{straizys} Straizys V, Multicolor Photometry of Stars [in Russian], Mokslas, Vilnius (1977).
\bibitem{strcor} Straizys V., Corbally C. J., and Laugalys V., Baltic astronomy 8, 355 (1999).
\bibitem{strub} Strub P., Kr\"uger H., and Sterken V. J., Astrophys. J. 812, 140 (2015).
\bibitem{struve} Struve F. G. W., Etudes d'Astronomie Stellaire, (Pulkovo observatory, 1847).
\bibitem{sumi} Sumi T., Mon. Not. Roy. Astron. Soc. 349, 193, (2004).
\bibitem{taylor} Taylor A. D., Baggaley W. J., Steel D. I., et al., Nature 380, 323 (1996).
\bibitem{thomas} Thomas P. C., Icarus, 208, 395 (2010).
\bibitem{vergely} Vergely J.-L., Freire Ferrero R., Egret D., et al., Astron. Astrophys. 340, 543 (1998).
\bibitem{vos} Voshchinnikov N. V., Journal of Quantitative Spectroscopy and Radiative Transfer 113, 2334 (2012).
\bibitem{wang2013} Wang S., Gao J., Jiang B. W., et al., Astrophys. J. 773, 30 (2013).
\bibitem{wang2014} Wang S., Li A., Jiang B. W., Planetary and Space Science, 100, 32 (2014).
\bibitem{wang2015} Wang S., Li A., Jiang B. W., Astrophys. J. 811, 38 (2015).
\bibitem{wegner} Wegner W., Astron. Nachr., 324, 219, (2003).
\bibitem{wd2001} Weingartner J. C., Draine B. T., Astrophys. J. 548, 296 (2001).
\bibitem{wesson} Wesson R., Barlow M. J., Matsuura M., et al., Mon. Not. Roy. Astron. Soc. 446, 2089 (2015).
\bibitem{witt} Witt A. N., Smith R. K., Dwek E., Astrophys. J. 550, L201 (2001).
\bibitem{wolf} Wolf C., Mon. Not. Roy. Astron. Soc. 445, 4252 (2014).
\bibitem{wise} Wright E. L., Eisenhardt P. R. M., Mainzer A. K. et. al., Astron. J. 140, 1868 (2010), http://irsa. ipac. caltech. edu/Missions/wise. html
\bibitem{yahata} Yahata K., Yonehara A., Suto Y., et al., Publ. Astron. Soc. Japan, 59, 205 (2007).
\bibitem{zasowski} Zasowski G., Majewski S. R., Indebetouw R., et al., Astrophys. J. 707, 510 (2009).
\bibitem{zemcov} Zemcov M., Smidt J., Arai T., et al., Science 346, 732 (2014).
\bibitem{zubko} Zubko V., Dwek E., Arendt R. G., Astrophys. J. Suppl. Ser. 152, 211 (2004).




\end{thebibliography}
\end{document}